\documentclass[preprint,1p,11pt,nonatbib]{elsarticle}
\makeatletter
\let\c@author\relax

\makeatother

\usepackage{amssymb}
\usepackage{amsmath}
\usepackage{amstext}
\usepackage{enumerate}
\usepackage{framed}
\usepackage{pdfpages}
\usepackage{graphicx}
\usepackage{diagbox}
\usepackage{caption,subfig}
\usepackage{booktabs}
\usepackage{lineno}
\usepackage{multirow}
\usepackage{url}
\usepackage{soul}
\setstcolor{red}
\sethlcolor{yellow}

\usepackage[
backend=biber,
style=numeric-comp,
sorting=none
]{biblatex}

\begin{document}
%\linenumbers
    \begin{frontmatter}
        \title{A New Optical Model for Photomultiplier Tubes}
        \author[ihep]{Yaoguang Wang}
        \author[ihep,ucas,keylab]{Guofu Cao\corref{cor}}
        \ead{caogf@ihep.ac.cn}
        \author[ihep,keylab]{Liangjian Wen}
        \author[ihep,keylab]{Yifang Wang}
        \cortext[cor]{Corresponding authors}
        \address[ihep]{Institute of High Energy Physics, Beijing 100049, China}
        \address[ucas]{University of Chinese Academy of Sciences, Beijing 100049, China}
        \address[keylab]{State Key Laboratory of Particle Detection and Electronics, Beijing 100049, China}
        \begin{abstract}
             It is critical to construct an accurate optical model of photomultiplier tubes (PMTs) in many applications to describe the angular and spectral responses of the photon detection efficiency (PDE) of the PMTs in their working media. In this study, we propose a new PMT optical model to describe both light interactions with the PMT window and optical processes inside PMTs with reasonable accuracy based on the optics theory and a GEANT4-based simulation toolkit. The proposed model builds a relationship between the PDE and the underlying processes that the PDE relies on. This model also provides a tool to transform the PDE measured in one working medium (like air) to the PDE in other media (like water, liquid scintillator, etc). Using two 20'' MCP-PMTs and one 20'' dynode PMT, we demonstrate a complete procedure to obtain the key parameters used in the model from experimental data, such as the optical properties of the antireflective coating and photocathode of the three PMTs. The proposed model can effectively reproduce the angular responses of the quantum efficiency of PMTs, even though an ideally uniform photocathode is assumed in the model. Interestingly, the proposed model predicts a similar level ($20\%\sim30\%$) of light yield excess observed in the experimental data of many liquid scintillator-based neutrino detectors, compared with that predicted at the stage of detector design. However, this excess has never been explained, and the proposed PMT model provides a good explanation for it, which highlights the imperfections of PMT models used in their detector simulations. 
        \end{abstract}
    \end{frontmatter}

    \section{Introduction}
    Photomultiplier tubes (PMTs) are common photosensors that are used extensively for weak light detection in a wide variety of applications, ranging from fundamental research to industrial applications. Typically, PMTs play an important role in large-scale neutrino projects, which typically use scintillators or ultrapure water as detector media and use PMTs to collect scintillation light and/or Cerenkov light. Although studies aimed at improving PMT performance have been performed for decades since the first PMT was successfully produced by Iams et al. in 1935~\cite{Iams:1935zz}, understanding the response of a PMT is still a critical and nontrivial task in many applications. Photon detection efficiency (PDE) is the most critical characteristic of a PMT. Knowledge of PMT PDE is an essential input to design a detector, to estimate its expected performance with reasonable accuracy, and to achieve a precise detector simulation that guarantees good control of relevant systematic errors and high-quality physics results. However, fully understanding a PDE is complex because PDEs strongly rely on both light interaction with the photocathode and optical processes inside PMTs, which will be discussed in detail in section~\ref{light_interaction}.   
    
    In the past, pioneering research primarily focused on light interactions with photocathodes. In~\cite{Moorhead:1992pt} and ~\cite{Moorhead:1991bc}, a major breakthrough was accomplished by Moorhead and Tanner, who proposed a method to infer the complex refractive index and thickness of the bialkali photocathode at a wavelength of 442~nm by measuring the angular dependence of the reflectance of an EMI 9124B PMT immersed in water. Later, more research was performed to investigate the optical properties of various photocathodes in a much wider spectral range, in which a model based on optics theory of thin films was also developed to describe the light interaction with the photocathode~\cite{Lay:97,Motta:2004yx,HARMER2006439,Matsuoka:2018oml}. Some studies also aimed to improve the quantum efficiency (QE) of PMTs using an antireflective coating (ARC) between the photocathode and PMT window, which enhances light absorption in the photocathode because of a better match of refractive indices~\cite{HALLENSLEBEN200089,Harmer_2012}. Even though it is sufficient to only consider the light interaction with photocathodes in some applications, it is essential to develop a comprehensive PMT optical model that accounts for all relevant optical processes, including those inside PMTs, due to its great importance in many applications, such as various PMT-based neutrino detectors. However, little information can be found regarding the development of such models. In this study, we propose a method to build a general PMT optical model that could be adapted to describe any type of PMT and is suitable for any external media. Due to the complexity of PMT optical processes, the new model is developed based on the optics theory of multilayer thin films, a simulation toolkit, and inputs from experimental measurements at a limited number of positions on PMTs. By considering three 20'' PMTs (including one dynode PMT manufactured by Hamamatsu and two MCP-PMTs manufactured by NNVT~\cite{mcppmtintro}) as an example, we demonstrate the procedure of building this model and extracting a few key parameters in the model. Interestingly, the proposed PMT model can provide a plausible explanation for the "light yield excess" that appears in many liquid scintillator and PMT-based neutrino detectors\mbox{~\cite{ALIMONTI2002205, Borexino:2008gab, DayaBay:2007fgu,  DayaBay:2012fng, Piepke:2001tg, KamLAND:2002uet, suekane2004overview, RENO:2010vlj, RENO:2012mkc}}. It shows that the light yield from experimental data is 20\%-30\% higher than that from simulation predictions at R\&D phases. This excess has never been explained, however, the proposed new model predicts a similar level of light yield excess. Therefore, we infer that this excess is caused by the imperfect PMT optical model used in their detector simulations, which is a simplified PMT model, assuming photons hitting on the photocathode are 100\% absorbed, then applying PDE to determine the number of detected photons.
    
    This paper is organized as follows. We first summarize the effects to be included in the new PMT optical model that PDE relies on. Then, a detailed description of the model is followed, including relevant optics theory of multilayer thin films and its integration with the GEANT4-based simulation toolkit~\cite{AGOSTINELLI2003250}. Next, we demonstrate the method to experimentally extract the key optical parameters in the model by measuring the angular dependence of the reflectance and QE of three 20'' PMTs. Finally, based on toy Monte Carlo simulation, comparisons of the light yields of a simple liquid scintillator detector are also presented between the proposed new PMT model and the simplified model, which highlight the potential reason for the aforementioned "light yield excess".
    
    \section{General consideration of PDE dependence}
    \label{light_interaction}
     The PDE is defined as the ratio of the number of detected photons to the number of incident photons on the PMT. In general, PDE can be further decomposed into QE and collection efficiency (CE). Based on Spicer's three-step model~\cite{PhysRev.112.114}, QE can be considered a product of two terms: one is the absorption probability of converting an incident photon to a photoelectron in the photocathode via the photoelectric effect, and the other is the escape probability of the generated photoelectrons that overcome the photocathode's potential and become free photoelectrons. CE is the probability of successfully collecting free photoelectrons via built-in electrodes of the PMT. It depends on the electrical field distributed inside the PMT. The electrical field accelerates the photoelectrons to hit the first dynode or micro-channel plate (MCP) and knock out secondary electrons. Then, the secondary electrons are multiplied in multistage dynodes (MCPs) and eventually collected by the anode to form an electrical signal. The following factors must be managed correctly to obtain an accurate PMT optical model aimed at describing the angular and spectral dependence of the PDE and its uniformity on PMTs.
     \begin{enumerate}[(1)]
         \item QE strongly depends on the wavelengths and angle of incidence (AOI) of incident photons.
         \item QE is dependent on the position of the PMT entrance window due to variations in photocathode properties caused by imperfect technologies and non-ideal environmental control during photocathode evaporation.
         \item The uniformity of the electrical field can be distorted by the shape and arrangement of the electrodes inside the PMT, which results in a position dependence of CE.
         \item A fraction of incident photons can pass through the PMT window and enter the vacuum region of the PMT; then, the transmitted photons can be reflected back to the photocathode and contribute to the QE and PDE. The contribution to the PDE from these photons varies with the AOI and wavelength of incident photons.
         \item QE is also influenced by the optical properties of the external medium, in which the PMT is operated. The external medium impacts the amplitudes of reflectance and transmittance at the interface between the external medium and the PMT window. The external medium also changes the AOI range of the incident photons and also the undergoing optical processes in the photocathode. For example, in Figure~\ref{fig:qe_plot} from Ref.~\cite{Motta:2004yx}, a significant difference in the PMT's QE is predicted between different external media of air and liquid scintillator, in particular, in the region of AOI larger than the critical angle, in which total internal reflection occurs only for the case of operating the PMT in liquid scintillator.
         \begin{figure}
         \centering
         \includegraphics[width=300px]{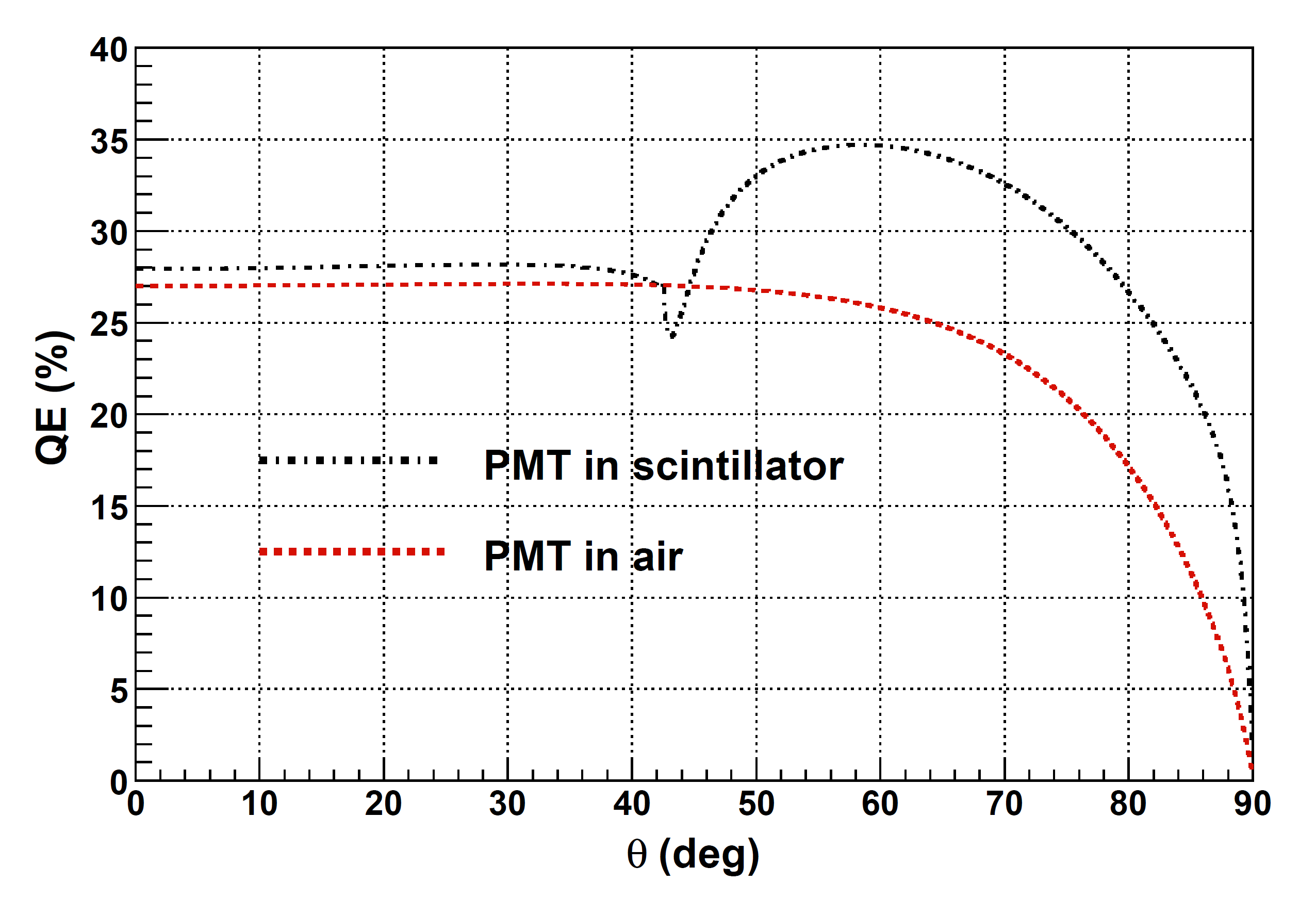}
         \caption{Predicted angular dependence of the QE, for the case of the PMT in liquid scintillator with a refractive index of 1.48 (black dashed-dotted curve) and in air (red dashed curve). The plot is taken from~\cite{Motta:2004yx}.}
         \label{fig:qe_plot}
         \end{figure}
         \item For a photodetector system consisting of multiple PMTs, reflected photons from one PMT might be captured again by other PMTs, which enhances the overall light collection efficiency of the photodetector system; thus, in this case, it is also an integral task to investigate the number of reflected photons from PMTs. In this work, we only present the results of PMTs' reflectance. Its impacts on the overall performance of a photodetector system are not yet adopted and should be considered in future studies.
     \end{enumerate}
     
     Although some aforementioned factors can be described by performing detailed characterizations of PMTs, a few factors are strongly coupled with each other. For example, the angular response of the PDE is correlated with the optical properties of external medium, as well as the optical processes occurred inside the PMT for some types of PMTs (e.g., the 20'' PMTs used in this study). The reason is that the measured PDE at a given position on the photocathode not only depends on the PDE at this position but also other positions due to the optical processes inside the PMT. Also, the contributions from other positions rely on the AOI of the incident photons, and different external media result in different AOIs on the photocathode. In addition, it is not realistic to perform all required evaluations for each PMT in some applications, which use a large number of PMTs, even up to a few thousand, such as JUNO ($\sim$20,000 20'' PMTs  and $\sim$25,000 3'' PMTs)~\cite{junocollaboration2021juno, JUNO:2015sjr} and Hyper-K ($\sim$40,000 20'' PMTs and $\sim$13,300 3'' PMTs)~\cite{Hyper-Kamiokande:2018ofw, HyperK-2019}. A PMT optical model is an ideal and effective approach to manage all of the aforementioned optical processes and precisely predict the PDE response of PMTs.
    
    \section{New PMT optical model}
    \subsection{General proposal of the model}
    In general, the PDE can be expressed as:
    \begin{equation}
        \mbox{PDE}(\lambda, \alpha)=\sum_{j}a_j(\lambda, \alpha)\cdot\rho_j(\lambda)\cdot \mbox{CE}_j = \sum_{j}a_j(\lambda,\alpha) F_j(\lambda)
        \label{eq:DE_Model}
    \end{equation}
    where $a_j$ and $\rho_j$ are the absorption ratio to the incident light and escape factor in the aforementioned three-step model at the $j$th position on the photocathode, respectively; $\lambda$ and $\alpha$ are the wavelength and AOI at the $j$th position, respectively. Because $\rho$ and CE are independent of AOI, we denote their product as a $F$-factor. The summation in Eq.~\ref{eq:DE_Model} indicates that the PDE might be contributed to by multiple other positions on the photocathode in addition to the incident position due to transmitted light transportation inside the PMT. The number of positions which contributes to the PDE not only depends on the wavelength and AOI of incident light, but also relies on the optical properties of inner structures inside the PMT. 
    
    Due to the complex optical processes inside the PMT, it is difficult to analytically calculate $a_j$ and $F_j$. However, $F_j$ can be interpolated with reasonable accuracy by a few selected reference positions on the photocathode. The accuracy of $F_j$ depends on the number and distribution of the reference positions, and on the uniformity of the photocathode. We assume that there are $n$ reference positions selected on the photocathode. At each reference position, its measured PDE is denoted as $\mbox{PDE}_i (i=1,2...n)$. According to Eq.~\ref{eq:DE_Model}, $\mbox{PDE}_i$ can be expressed as:
    \begin{equation}
        \mbox{PDE}_i = \sum_j a_{ij}F'_j = \sum_j a_{ij}\sum_{k=1}^{n}\beta_{jk}F_k
        \label{eq:DE_i}
    \end{equation}
    where the $F$-factors at each contributed position are denoted as $F'_j$, which can be interpolated with the value of $F_k$ and its weight coefficient, represented as $\beta_{jk}$. $F_k$ denotes the $F$-factor at the $k$th reference position. Eq.~\ref{eq:DE_i} is valid for any AOIs and any positions on the photocathode in addition to the reference positions. This equation can also be written as:
    \begin{equation}
        \mbox{PDE}_i = \sum_j a_{ij} \sum_{k=1}^n \beta_{jk}F_k = \sum_{k=1}^n\sum_j a_{ij}\beta_{jk}F_k = \sum_{k=1}^n A_{ik} F_k
        \label{eq:DE_F_relation}
    \end{equation}
    where $A_{ik}=\sum_{j}a_{ij}\beta_{jk}$. Eq.~\ref{eq:DE_F_relation} can also be rewritten in a matrix form:
    \begin{equation}
        \begin{pmatrix}
            \mbox{PDE}_1 \\
            \mbox{PDE}_2 \\
            \vdots \\
            \mbox{PDE}_n
        \end{pmatrix}
        =
        \begin{pmatrix}
            A_{11} & \cdots & A_{1n} \\
            \vdots & \ddots & \vdots \\
            A_{n1} & \cdots & A_{nn}
        \end{pmatrix}
        \begin{pmatrix}
            F_1 \\
            F_2 \\
            \vdots \\
            F_n
        \end{pmatrix}
        \label{eq:DE_F_matrix}
    \end{equation}
    It is easy to determine the PDE matrix on the left-hand side of Eq.~\ref{eq:DE_F_matrix} by performing the PDE measurements at the reference positions with a fixed AOI. The value of the AOI can be chosen according to convenience during the PDE measurements because it only impacts the $A$ matrix, and the $F$ matrix is independent of the AOI. Once the $A$ matrix is known, the $F$ matrix can be resolved analytically. 
    
    The elements in the $A$ matrix include contributions from the absorption of the photocathode at different positions and the weight coefficient of each position that is determined by the optical processes inside the PMT. Both the absorption coefficient and weight coefficient can be obtained by combining a general Monte Carlo simulation toolkit, such as GEANT4, and optics theory of multilayer thin films, which will be discussed in section~\ref{optics_theory}. However, in this process, a few key steps are required to be addressed ahead. The optical properties of external media, PMT windows, ARCs (if existing) and photocathodes must be known first, where the optical constants of the ARC and photocathode are the most critical and usually remain confidential. In section~\ref{opticsparam}, we demonstrate how to obtain this information for three 20'' PMTs using the method proposed by Moorhead and Tanner. Regarding other optical parameters, it is not difficult to find most in the literature. Then, a detailed PMT geometry model must be carefully constructed because it is an essential input to simulate optical processes inside the PMT with good accuracy, which leads to a better estimation of the $A$ matrix. The geometric model should include the PMT bulb and inner components with their optical properties, such as various electrodes, reflective aluminum films, and supporting structures. In the end, it is straightforward to integrate optics theory of multilayer thin films with the simulation toolkit if this functionality is not yet implemented, such as GEANT4 used in this study. Then, the simulation toolkit can manage light interactions with the photocathode and complete simulations of various optical processes, including those occurring inside the PMT, with the known optical properties of relevant materials. Eventually, the $A$ matrix can be determined from the simulation, and the $F$ matrix can be resolved. Once the $F$ matrix is resolved, it can be used together with the $A$ matrix to predict the PDE for any AOI at any position and in any external media. The $F$-matrix is a function of wavelength because it includes the contribution from the escape factor, which is spectrally dependent. Therefore, the aforementioned processes must be performed at different wavelengths to obtain the spectral response of the $F$-matrix. Eq.~\ref{eq:DE_F_matrix} is also suitable for the case of simply replacing PDE by QE; however, in this case, the $F$-matrix no longer relies on CE. Then, following the same procedure discussed above, the $F$-matrix can be obtained, and QE can be predicted. 
    
    Assuming that both photocathode properties and CE are ideally uniform on the PMT entrance window, which is a good approximation in some applications, Eq.~\ref{eq:DE_Model} reduces to a simple equation:
    \begin{equation}
        \mbox{PDE}(\lambda, \alpha) = F(\lambda)\cdot \sum_j a_{j}(\lambda, \alpha) = \rho(\lambda)\cdot \mbox{CE}\cdot \sum_j a_{j}(\lambda, \alpha)
        \label{eq:PDE_simple}
    \end{equation}
    Then, with the known $a_j$ determined by Monte Carlo simulation, the $F$ value can be estimated by measuring the PDE at one position on the photocathode. Again, Eq.~\ref{eq:PDE_simple} can be directly applied for QE by setting the CE to be 1. If the contribution to the PDE is negligible from the optical processes inside the PMT, which is the case for some types of PMTs, Eq.~\ref{eq:PDE_simple} can be further simplified into the well-known form of:
    \begin{equation}
        \mbox{PDE}(\lambda, \alpha) = F(\lambda)\cdot a(\lambda, \alpha) = \rho(\lambda)\cdot \mbox{CE}\cdot a(\lambda, \alpha) = \mbox{QE}(\lambda, \alpha)\cdot \mbox{CE}
        \label{eq:PDE_simple_noreflection}
    \end{equation}
    where QE is a product of the absorption factor and the escape factor.
    
    When we consider a photodetector system, reflected photons from a PMT can significantly impact its performance in some applications, such as JUNO, in which approximately 18,000 20'' PMTs are deployed on a sphere with a photocathode coverage as high as $\sim$75\%~\cite{JUNO:2015sjr}. The reflected photons have a large probability of being detected again by other PMTs in the system. These photons contribute to light yield and affect the hit pattern and hit time distribution, which are important for reconstruction and particle identification. The amount of reflected photons from a PMT consists of two components: the photons reflected at the entrance position, whose reflectivity can be easily calculated via optics theory of multilayer thin films with known optical constants of external medium, PMT window and photocathode; and the photons that transmit into the PMT and are reflected back to the PMT window by the inner structures and then exit into the external medium. In this case, a reliable simulation toolkit and a detailed PMT geometry model can manage the complex optical processes inside the PMT. Light interactions with photocathodes and PMT windows can always be predicted by optics theory.

    \subsection{Optics theory of multilayer thin film}
    \label{optics_theory}
    In general, the structure of the PMT window at the incident point can be regarded as a stack of several planar layers, which is also a good approximation, even for a bulb-like PMT, because of much shorter wavelengths of several hundred nanometers of interest, compared with the radii of curvature of PMTs' bulbs. Therefore, from outside to inside, the stack consists of an external medium, PMT glass, ARC, photocathode and PMT vacuum, as shown in Figure~\mbox{\ref{fig:PMT_stack}}. ARC and photocathode are the so-called coherent layers, which means that their thicknesses are comparable to the wavelengths of incident light and that the reflected (transmitted) waves interfere with each other. Other layers are incoherent since their thicknesses are much larger than the wavelengths, and the reflected (transmitted) waves lose their coherence. We now briefly summarize the relevant optics theory used in this study.
    
    Reflectance ($R$) and transmittance ($T$) are defined as the ratio of reflected or transmitted irradiance to incident irradiance. For a simple boundary between the two media, they can be denoted as:
    \begin{align}
        R&=|r|^2 \notag \\
        T&=\frac{n_2\cos{\theta_2}}{n_1\cos{\theta_1}}|t|^2
        \label{eq:RT}
    \end{align}
    where $r$ and $t$ are the reflection and transmission amplitudes from the first medium (medium 1) to the second medium (medium 2), respectively, and they can be calculated using Fresnel's equations:
    \begin{align}
        r_s &= \frac{n_1\cos{\theta_1}-n_2\cos{\theta_2}}{n_1\cos{\theta_1}+n_2\cos{\theta_2}}, \quad t_s= \frac{2n_1\cos{\theta_1}}{n_1\cos{\theta_1}+n_2\cos{\theta_2}} \notag \\
        r_p &= \frac{n_2\cos{\theta_1}-n_1\cos{\theta_2}}{n_2\cos{\theta_1}+n_1\cos{\theta_2}}, \quad t_p= \frac{2n_1\cos{\theta_1}}{n_2\cos{\theta_1}+n_1\cos{\theta_2}}
        \label{eq:rt}
    \end{align}
    The subscripts $s$ and $p$ identify the s-polarized and p-polarized light, respectively, and $n_1$ and $n_2$ are the refractive indices of medium 1 and medium 2, respectively. $\theta_1$ and $\theta_2$ are angles of incidence and refraction that follow Snell's law:
    \begin{equation}
        n_1\sin{\theta_1} = n_2\sin{\theta_2}
    \end{equation}
    Therefore, for this simple boundary, such as that between the two incoherent layers of the external medium and PMT glass, $R$ and $T$ at any AOI can be easily calculated with the known refractive indices of the two media. These formulae are still valid if the light absorption in the two media is not negligible; however, in this case, the refractive indices become complex numbers, and the corresponding amplitudes of reflection and transmission are also complex values.
    
    Analysis becomes marginally more complex when we consider a stack consisting of multiple coherent or incoherent layers, in which light can be reflected multiple times between two adjacent boundaries. The transfer matrix method is a general approach to manage both multicoherent and multi-incoherent layer systems~\cite{tmm, byrnes2020multilayer}. We do not report details about deriving this method and only summarize the final expressions used in this study. We assume that there are $N$ layers numbered with $1\cdots N$ arranged between medium $0$ and medium $N+1$. According to the properties of the $N$ layers, the following three cases are considered, including $N$ coherent layers, $N$ incoherent layers, and the hybrid structure.
    
    \subsubsection{$N$ coherent layers}
\begin{figure}
        \centering
        \subfloat[]{
        \label{fig:PMT_stack}
        \includegraphics[scale=1.0]{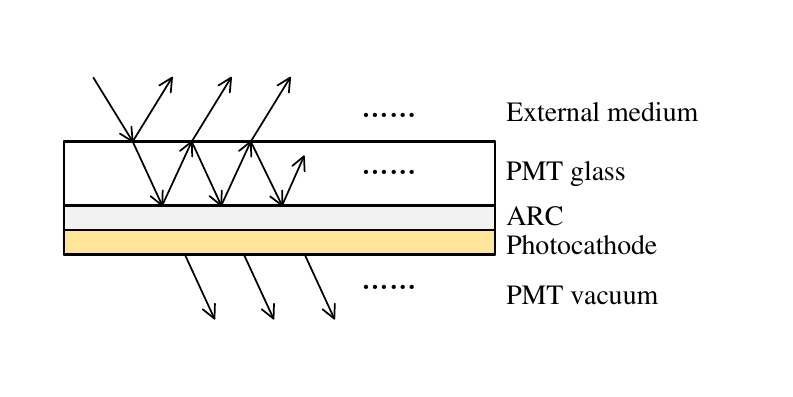}
        }
        \subfloat[]{
        \label{fig:TMM_sketch}
        \includegraphics[scale=0.35]{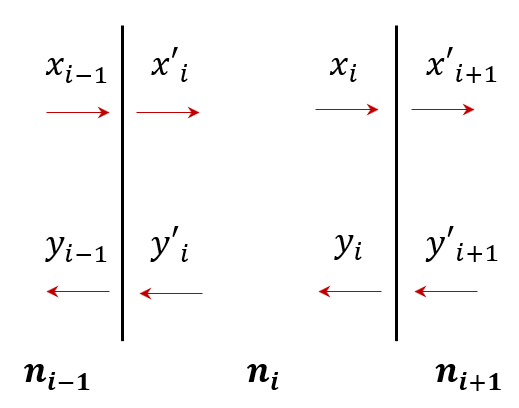}
        }
        \caption{Light reflection, refraction and propagation in the stack. Figure~\mbox{\ref{fig:PMT_stack}} shows the stack structure of the PMT window. Figure~\mbox{\ref{fig:TMM_sketch}} stands for the general case of a multilayer system.}
    \end{figure}

    If all of the $N$ layers are coherent, such as ARC and photocathode, it is necessary to calculate the reflection and the transmission in view of field amplitudes. Figure~\ref{fig:TMM_sketch} shows the notations of field amplitudes. We denote the amplitude of the right-going (left-going) light with $x$ ($y$); the subscripts indicate which layer the light lies in, and the prime implies that the light is at the left boundary of the medium. It is easy to identify the relationships of the field amplitudes at each boundary:
    \begin{equation}
        \begin{pmatrix}
            x_{i-1} \\
            y_{i-1}
        \end{pmatrix}
        =\frac{1}{t_{i-1,i}}
        \begin{pmatrix}
            1 & r_{i-1, i} \\
            r_{i-1, i} & 1
        \end{pmatrix}
        \begin{pmatrix}
            x'_i \\
            y'_i
        \end{pmatrix}
        =m_{i-1 \to i} 
        \begin{pmatrix}
            x'_i \\
            y'_i 
        \end{pmatrix}
        \label{eq:transmission_matrix}
    \end{equation}
    where $m_{i-1 \to i}$ is called the transmission matrix. Conversely, the phase of light changes during its propagation in the medium, which can be expressed with the propagation matrix:
    \begin{equation}
        \begin{pmatrix}
            x'_i \\
            y'_i
        \end{pmatrix}
        =\begin{pmatrix}
            e^{-i\delta_i} & 0 \\
            0 & e^{i\delta_i} 
        \end{pmatrix}
        \begin{pmatrix}
            x_i \\
            y_i
        \end{pmatrix}
        =p_i
        \begin{pmatrix}
            x_i \\
            y_i
        \end{pmatrix}
        \label{eq:propagation_matrix}
    \end{equation}
    where $\delta_i=2\pi n_i d_i \cos\theta_i /\lambda$. Left-going light vanishes in medium $N+1$; therefore, $y'_{N+1}=0$. Thus, with Eq.~\ref{eq:transmission_matrix} and Eq.~\ref{eq:propagation_matrix}, the relationships of the incident, reflected and transmitted light of the stack structure can be obtained:
    \begin{align}
        \begin{pmatrix}
            x_0 \\
            y_0
        \end{pmatrix}
        &=m_{0 \to 1} \prod_{i=1}^{N}p_i m_{i \to i+1}
        \begin{pmatrix}
            x'_{N+1} \\
            0
        \end{pmatrix} \nonumber \\ 
        &=\begin{pmatrix}
            \tilde{m}_{00} & \tilde{m}_{01} \\
            \tilde{m}_{10} & \tilde{m}_{11}
        \end{pmatrix}
        \begin{pmatrix}
            x'_{N+1} \\
            0
        \end{pmatrix}
        =\tilde{m}
        \begin{pmatrix}
            x'_{N+1} \\
            0
        \end{pmatrix}
    \end{align}
    The corresponding reflection and transmission amplitudes can be expressed with the matrix elements of $\tilde{m}$:
    \begin{align}
        r&=\frac{y_0}{x_0}=\frac{\tilde{m}_{10}}{\tilde{m}_{00}} \\ 
        t&=\frac{x'_{N+1}}{x_0} = \frac{1}{\tilde{m}_{00}}
    \end{align}
    Combined with Eq. \ref{eq:RT}, the reflectance and transmittance of the multi-coherent layers can be obtained.
    
    \subsubsection{$N$ incoherent layers}
    For the incoherent case, it is convenient to calculate the reflection and the transmission in view of power, and we use capital letters to denote the power of light beams. Similarly, we can also introduce the transmission matrix $M$:
    \begin{equation}
        \begin{pmatrix}
            X_{i-1} \\
            Y_{i-1}
        \end{pmatrix}
        =\frac{1}{T_{i-1, i}}
        \begin{pmatrix}
            1 & -R_{i, i-1} \\
            R_{i-1, i} & T_{i, i-1}T_{i-1, i}-R_{i,i-1}R_{i-1, i}
        \end{pmatrix}
        \begin{pmatrix}
            X'_{i} \\
            Y'_{i}
        \end{pmatrix}
        =M_{i-1 \to i}
        \begin{pmatrix}
            X'_i \\
            Y'_i
        \end{pmatrix}
    \end{equation}
    and the propagation matrix $P$:
    \begin{equation}
        \begin{pmatrix}
            X'_i \\
            Y'_i
        \end{pmatrix}
        =
        \begin{pmatrix}
            e^{\Delta_i} & 0 \\
            0 & e^{-\Delta_i}
        \end{pmatrix}
        \begin{pmatrix}
            X_i \\
            Y_i
        \end{pmatrix}
        =P_i
        \begin{pmatrix}
            X_i \\
            Y_i
        \end{pmatrix}
    \end{equation}
    where $\Delta_i=4\pi d_i \text{Im}[n_i \cos\theta_i]/\lambda$. Finally, we can obtain the relationships that describe the power of light beams:
    \begin{align}
        \begin{pmatrix}
            X_0 \\
            Y_0
        \end{pmatrix}
        &=M_{0 \to 1}\prod_{i=1}^{N}P_i M_{i \to i+1}
        \begin{pmatrix}
            X'_{N+1} \\
            0
        \end{pmatrix} \nonumber \\
        &=
        \begin{pmatrix}
            \tilde{M}_{00} & \tilde{M}_{01} \\
            \tilde{M}_{10} & \tilde{M}_{11}
        \end{pmatrix}
        \begin{pmatrix}
            X'_{N+1} \\
            0
        \end{pmatrix}
        =\tilde{M}
        \begin{pmatrix}
            X'_{N+1} \\
            0
        \end{pmatrix}
         \label{eq:TMM_Incoherent}
    \end{align}
    The reflectance and transmittance can be directly obtained with the matrix elements of $\tilde{M}$:
    \begin{align}
        R &= \frac{Y_0}{X_0} = \frac{\tilde{M}_{10}}{\tilde{M}_{00}} \label{opt_fit1}\\
        T &= \frac{X'_{N+1}}{X_0}=\frac{1}{\tilde{M}_{00}}
    \end{align}
    
    \subsubsection{Hybrid structure}
    For a hybrid structure that consists of both coherent and incoherent layers, we can select the incoherent layers and number them from $1$ to $N$ in order; then, Eq.~\ref{eq:TMM_Incoherent} is still valid. If coherent layers exist between medium $i$ and medium $i+1$, $R$ and $T$ in $M_{i \to i+1}$ should be the total reflectance and transmittance of the multi-coherent layers, as discussed above. In particular, we consider the PMT window, which contains one incoherent layer (PMT glass) and two coherent layers (ARC and photocathode) between the outside medium and PMT vacuum. The $\tilde{M}$ matrix in this case can be denoted by:
    \begin{equation}
        \tilde{M} = M_{\text{out}\to\text{glass}} P_{\text{glass}} M_{\text{glass}\to\text{vacuum}}
    \label{opt_fit2}
    \end{equation}
    
    $R$ and $T$ on the PMT window can be accurately calculated with the known optical properties (refractive index, extinction coefficient and thickness) of each layer in the stack structure. Conversely, the optical parameters of each layer can also be extracted using the formulae above to fit experimental reflectance or transmittance data in some liquids because the aforementioned feature of total internal reflection imposes strong constraints on these optical parameters, and this feature does not exist in experimental data collected in air.
    
    \subsection{GEANT4 improvements and PMT geometry model}
    \label{pmtgeometry}
    In GEANT4, Fresnel's equations are used to calculate the probabilities of light reflection and refraction at the boundary between two adjacent media, and users must provide information on the refractive indices of the two media. However, it cannot handle optical processes in coherent layers where light interference occurs, such as ARCs and photocathodes. In this study, we improve GEANT4 functionality by implementing the optics theory of multilayer thin films. A new package based on transfer matrix method is developed first to calculate the reflectance, transmittance and absorbance of a stack structure that consists of multiple thin film layers with an arbitrary number. Then, we define the ARC and photocathode as an optical surface between the PMT glass and vacuum in GEANT4, and relevant optical parameters are stored in its material property table. Finally, we use a user-defined physics process in GEANT4 that is implemented via the fast simulation method. The fast simulation is triggered only when light strikes the predefined optical surface. Once it is triggered, the fast simulation model takes over the optical processes and uses the new developed package to calculate the reflectance, transmittance and absorbance of the stack structure based on optical parameters of the stack structure and wavelengths; AOIs; and the polarization of incident light. For a photon absorbed by the photocathode, the $F$ factor defined in Eq.~\ref{eq:DE_F_matrix} is used to determine whether light is detected or not. Regarding reflected or transmitted light, GEANT4 will automatically take over and complete the simulation of the remaining optical processes. 
    
\begin{figure}
        \centering
        \subfloat[MCP-PMT]{
        \label{fig:fig1a}
        \includegraphics[scale=0.3]{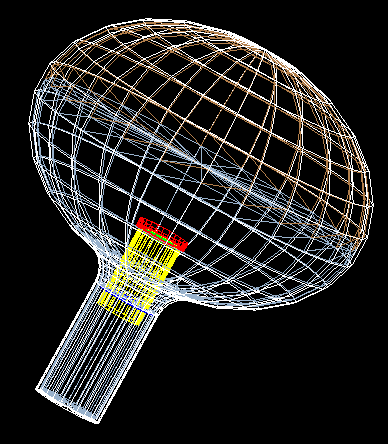}
        }
        \subfloat[Dynode-PMT]{
        \label{fig:fig1b}
        \includegraphics[scale=0.325]{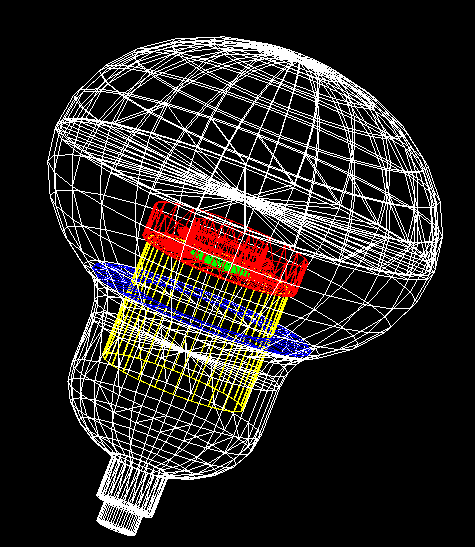}}
        \caption{Geometries of two types of 20'' PMTs implemented based on GEANT4. Fig~\ref{fig:fig1a} is the MCP-PMT manufactured by NNVT, and Fig~\ref{fig:fig1b} is the dynode PMT manufactured by Hamamatsu. }
        \label{fig:fig1}
    \end{figure}

    Detailed PMT geometry models are constructed with tools provided in GEANT4 and shown in Figure~\ref{fig:fig1} for the two types of 20'' PMTs: the left PMT is a 20'' MCP-PMT manufactured by NNVT, and the right one is a 20'' dynode PMT produced by Hamamatsu. With dimensions of the PMT bulbs listed in their datasheet~\cite{datasheet}, the PMT windows are defined as ellipsoids, the semimajor and semiminor axes of which are 254~mm and 190~mm for the dynode PMT (254~mm and 184~mm for the MCP-PMT), respectively, and the average thickness of the PMT glass is determined to be approximately 4~mm by an ultrasonic thickness gauge. The inner surfaces of the upper semispheres are defined as optical surfaces that represent the ARC and photocathode, and the lower semispheres evaporated with aluminum are also defined in the same way. For the inner components, it is difficult to consider all the details in the simulation because some fine structures exist; thus, some simplifications are made. Basically, the inner components consist of three parts: a cylindrical tube at the bottom (yellow), a focusing electrode on the top (red), and a dynode (MCP) (green) in the center of the focusing electrode. The dimensions of inner components are from a rough estimation, since little information about them can be found in the literature. By varying the dimensions of the inner structure with 10\% in the model, we found that its impact on the angular response of QE is not significant.
    
    \section{Extraction of optical parameters in the PMT optical model}
    \label{opticsparam}
\begin{figure}
        \centering
        \includegraphics[width=300px]{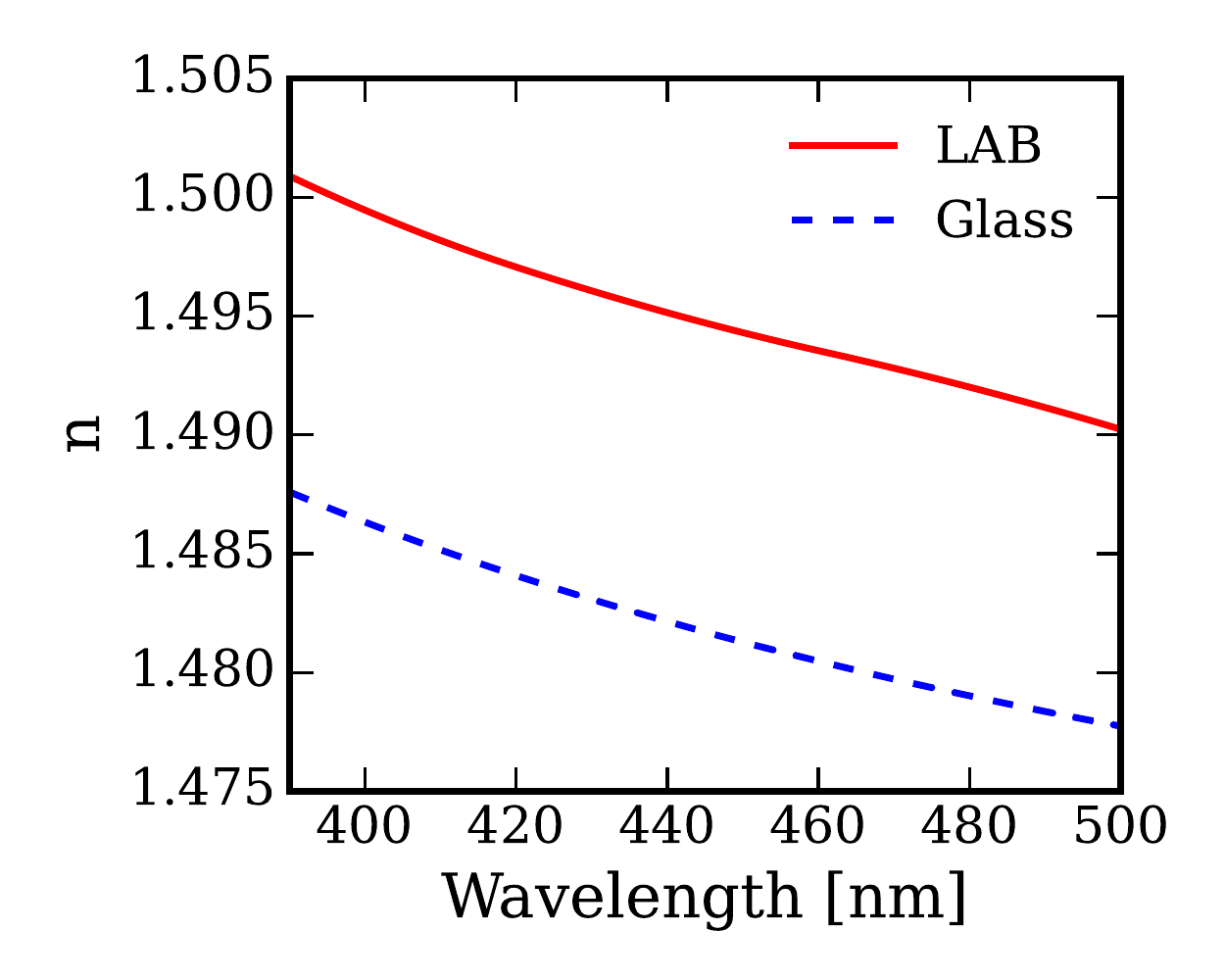}
        \caption{Refractive indices of LAB and PMT glass}
        \label{fig:nk_glass_lab}
    \end{figure}

    Refractive indices, extinction coefficients and thicknesses of ARC and photocathode are the key optical parameters in the new PMT optical model, in which the first two items are spectrally dependent. These parameters can be extracted using Eq.~\ref{opt_fit1} and Eq.~\ref{opt_fit2} to fit the angular response of the PMT reflectance in liquids. In this study, we consider three 20'' PMTs as an example to demonstrate the procedure of obtaining the optical parameters of interest; two are 20'' MCP-PMTs, where one is labelled a normal-QE tube, and the other is called a high-QE tube with improved technologies during photocathode fabrication. The third PMT is a 20'' dynode PMT (model number R12860). We measure the reflectance in the linear alkylbenzene (LAB) for the three PMTs because the LAB has a good index match with the PMT glass, as shown in Figure~\ref{fig:nk_glass_lab} from measurements. The reflectance at the interface between LAB and the PMT glass is calculated to be far less than 1\%, which is considered to be a systematic error in the reflectance of the ARC and photocathode in LAB. 
    
    \subsection{Experimental setup, measurement principle and uncertainties}
\begin{figure}
        \centering
        \includegraphics[scale=0.6]{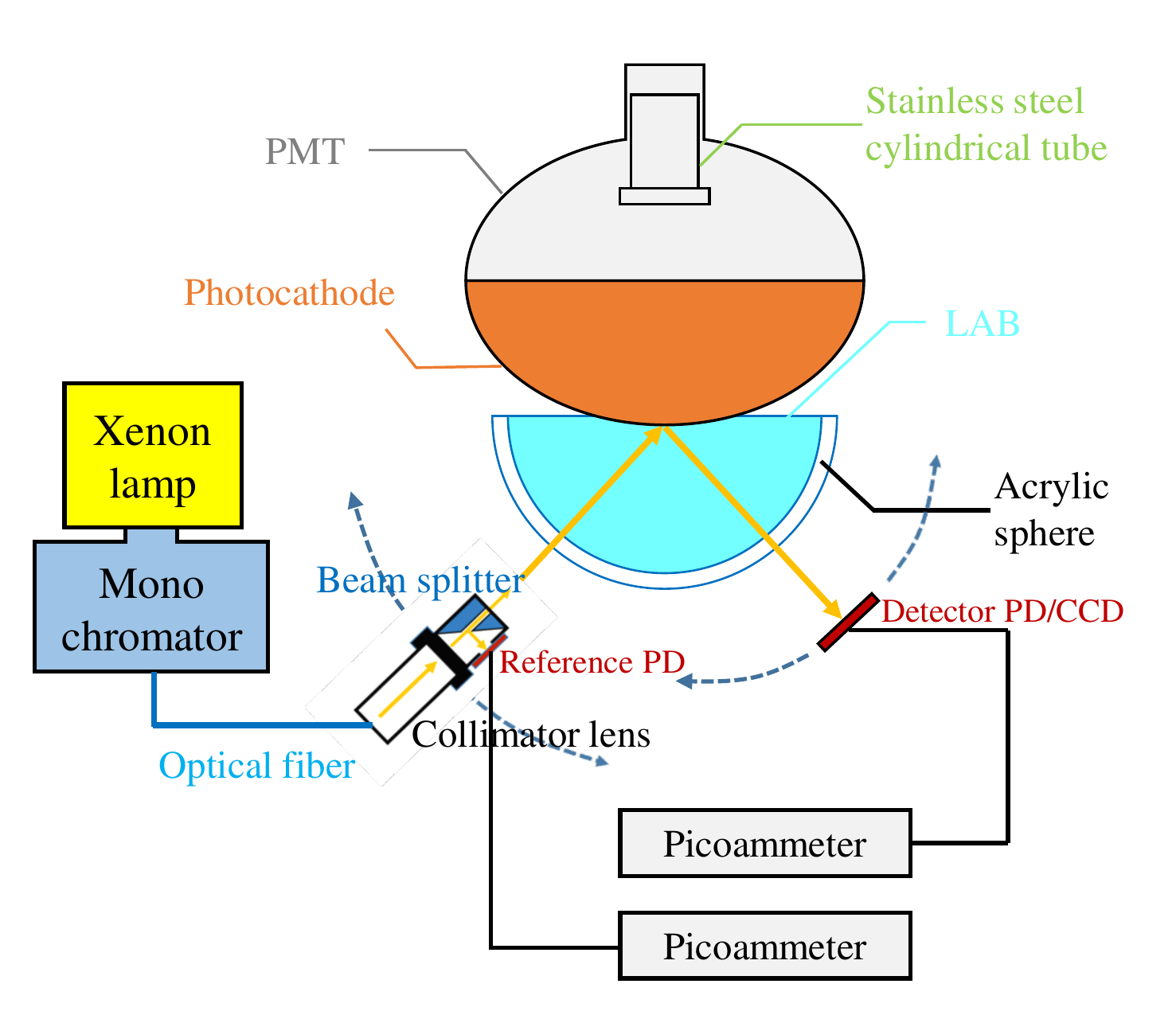}
        \caption{Schematic diagram of the experimental setup}
        \label{fig:exp_setup}
    \end{figure}

    The schematic diagram in Figure~\ref{fig:exp_setup} shows the experimental setup used to measure the PMTs' reflectance in LAB. A detailed description of this setup can be found in Ref.~\cite{Wang:2020mdl}. A xenon lamp serves as a light source to produce continuous and unpolarized light with wavelengths from $\sim$300 to $\sim$1200~nm. The lamp is coupled to a monochromator for wavelength selection with a resolution of approximately 2~nm. An optical fiber guides the light from the monochromator to a collimating lens located in a dark room. Then, a custom-made beam splitter divides the collimated light into two beams. The light intensity of one beam is monitored by a reference photodiode (PD), and another beam illuminates the PMT being tested. Due to the large dimensions of the 20'' PMTs, only the measured point on the PMT is immersed in LAB contained in an acrylic semisphere. The position of the measured point is carefully adjusted to ensure that it is at the center of the acrylic semisphere. The profile of the incident light beam is measured with a CCD camera, which has a diameter of 3~mm. The reflected light from the PMT is detected by another PD, which is called the detector PD. The currents of both PDs are simultaneously recorded by the two picoammeters, which are proportional to the light intensity of the light beams illuminated on the PDs. The collimator lens, beam splitter and reference PD are mounted on one rotary arm, which is controlled by a step motor to select AOIs. The detector PD is deployed on another rotary arm that is used to scan and find the maximum intensity of the reflected light beam. The minimum AOI can measure 15 degrees of reflectance because interference exists between the two rotary arms but can start from 0 degrees in the PMTs' QE evaluations in LAB using the same setup to measure the cathode current, in which a sufficiently high voltage is applied between the cathode and the dynode (or MCP).
    
    We follow the same analysis procedures discussed in Ref.~\cite{Wang:2020mdl} to obtain the results of reflectance in LAB and its systematic uncertainties. The stability of light intensity and the uniformity of the acrylic semisphere are the two major factors contributing to systematic errors. They are estimated to be 0.98\% and 1\%, respectively, which yield final uncertainties of the reflectance in LAB of 3\% after error propagation. The non-uniformity of the semisphere is caused by the variation of the thickness and the defects on the surface during its production. Both the reflectance data and QE data in the LAB of the three 20'' PMTs are collected at wavelengths from 390 to 500~nm in steps of 10~nm. At each wavelength, the AOI is scanned from 15 (0 for QE data) to 70 degrees in steps of 1 degree around the region of the critical angle of the total internal reflection and 2 degrees for other regions. 
    
    \subsection{Optical parameters of ARC and photocathode}
\begin{figure}
        \centering
        \subfloat[$\lambda=400~\mbox{nm}$]{
        \label{fig:fig3a}
        \includegraphics[scale=0.55]{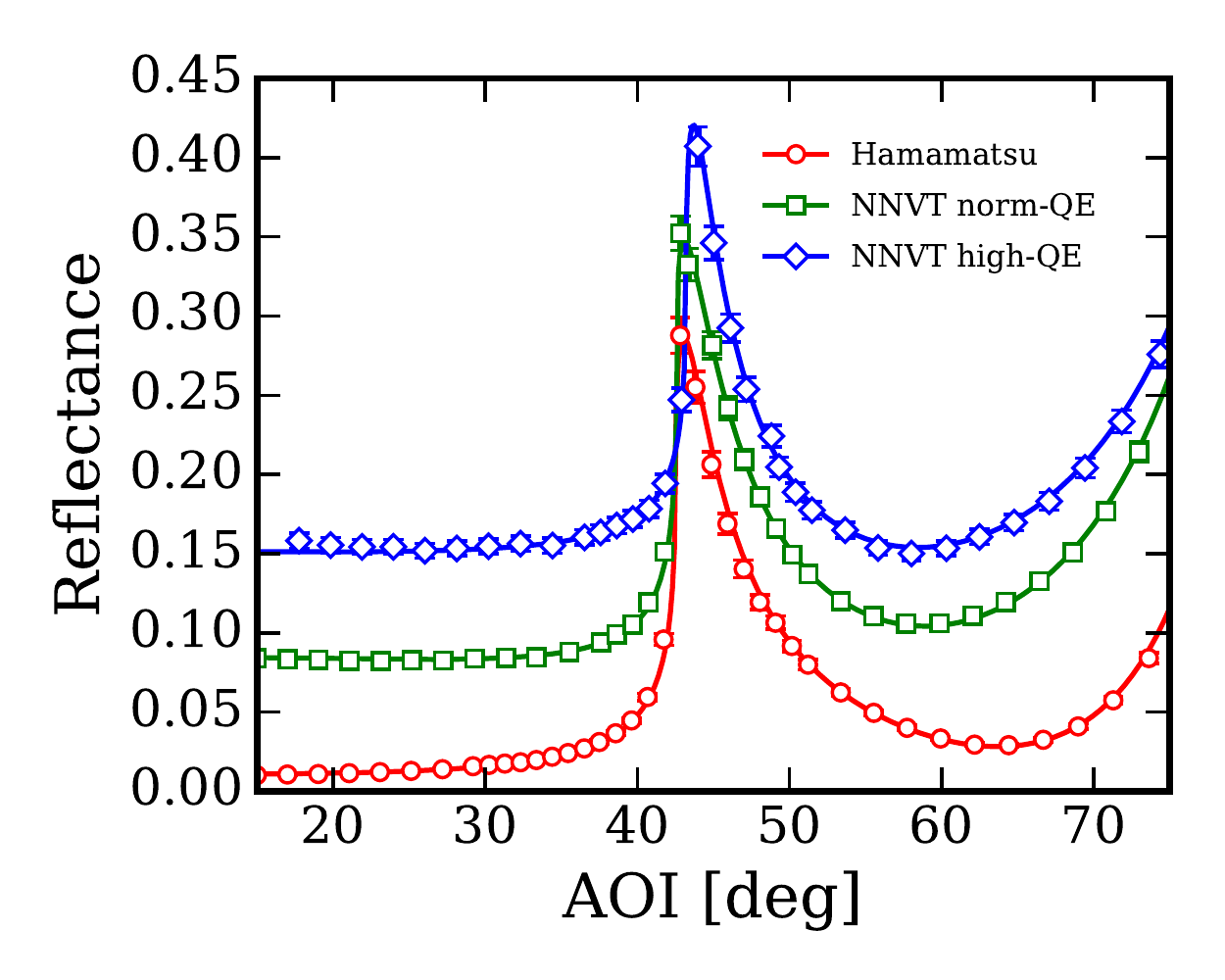}
        }
        \subfloat[$\lambda=420~\mbox{nm}$]{
        \label{fig:fig3b}
        \includegraphics[scale=0.55]{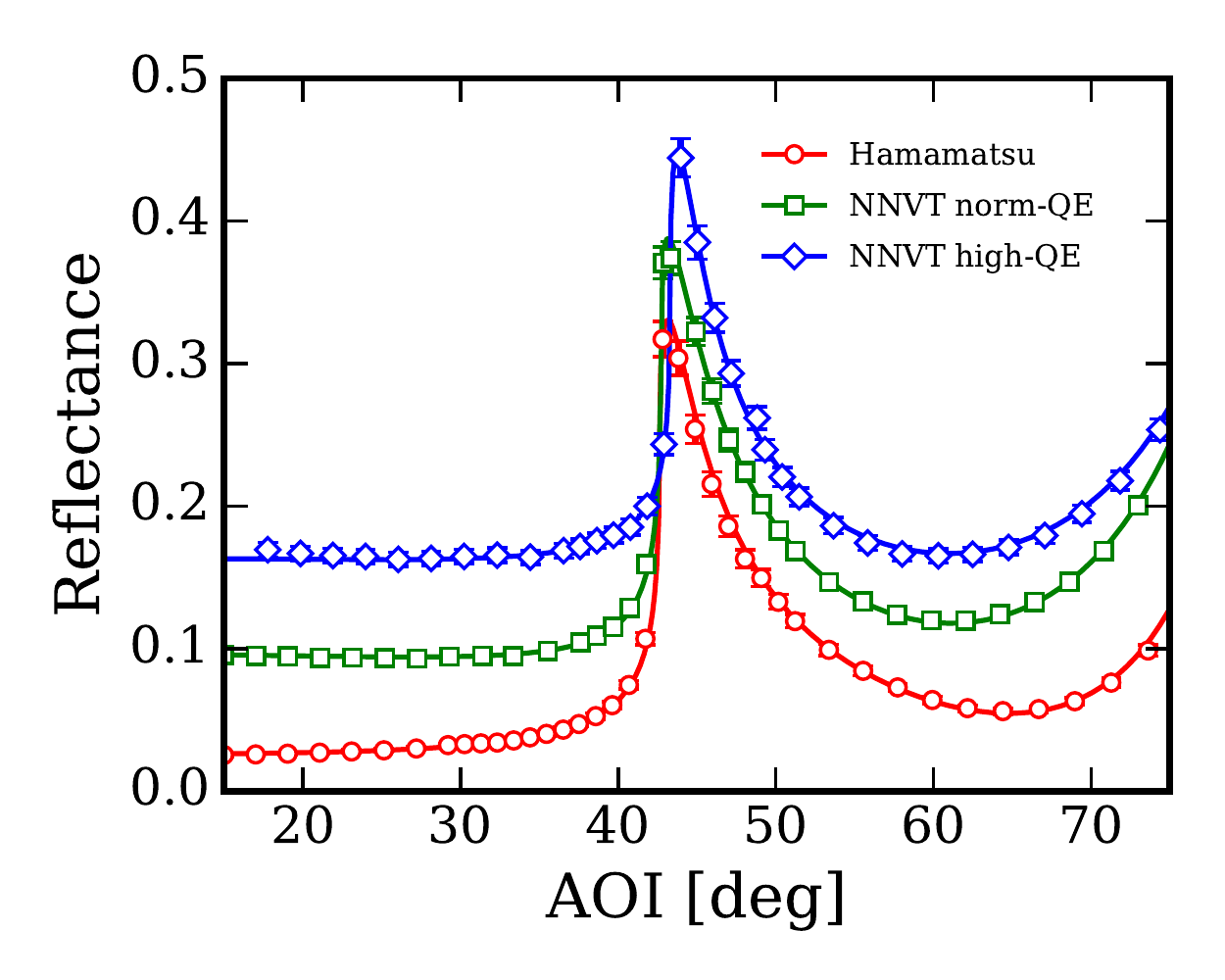}
        }
        
        \subfloat[$\lambda=450~\mbox{nm}$]{
        \label{fig:fig3c}
        \includegraphics[scale=0.55]{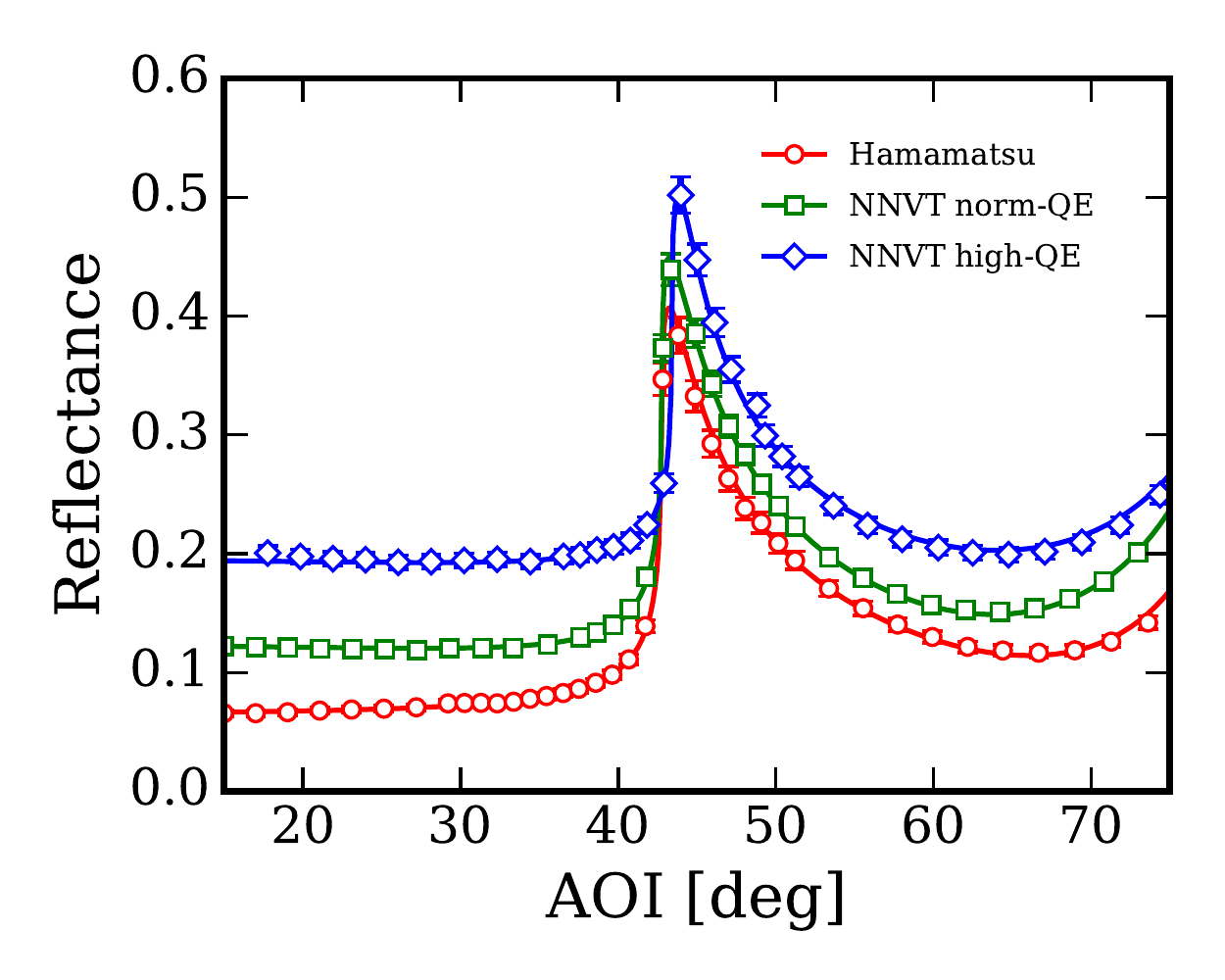}
        }
        \subfloat[$\lambda=500~\mbox{nm}$]{
        \label{fig:fig3d}
        \includegraphics[scale=0.55]{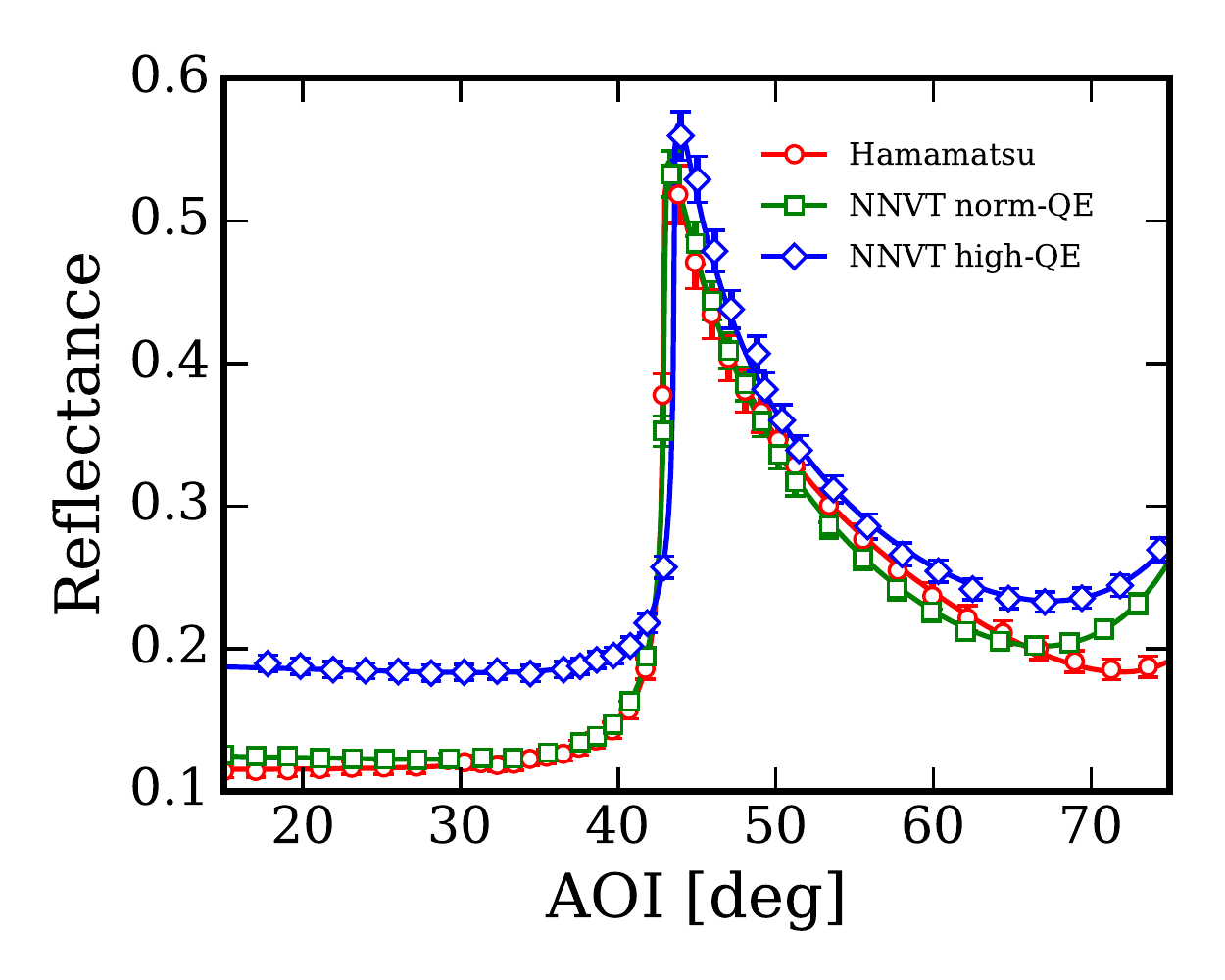}
        }
        \caption{Reflectance in LAB as a function of AOI, measured for R12860 (red), normal-QE MCP-PMT (green) and high-QE MCP-PMT (blue) at four selected wavelengths of 400~nm (a), 420~nm (b), 450~nm (c) and 500~nm (d).}
        \label{fig:fig3}
    \end{figure}

    Figure~\ref{fig:fig3} shows the typical angular response of reflectance in LAB for the three different PMTs at four selected wavelengths: 400~nm, 420~nm, 450~nm and 500~nm. The peaks at approximately 42 degrees are caused by the total internal reflection at the boundary between the photocathode and the vacuum when the PMTs are immersed in LAB. This feature is important to constrain the optical parameters of the ARC and photocathode but cannot be observed when the PMTs are characterized in air. The different reflectance values of the three PMTs indicate the different materials or technologies used in the fabrication of ARCs and photocathodes.
   
    Since reflectance data are collected at a fixed position on each measured PMT, the thicknesses of the ARC and photocathode remain the same for the different wavelengths. Therefore, a combined fit is performed to fit the angular response of the reflectance for all measured wavelengths using Eq.~\ref{eq:TMM_Incoherent}. A $\chi^2$ function is constructed and minimized to obtain the best-fit values of optical parameters and estimate their uncertainties. Eq.~\ref{eq:chi2} defines the $\chi^2$ function:
    \begin{equation}
        \chi^2 = \sum_i\sum_j\left(\frac{R_{theo}(\lambda_i, \alpha_j, \delta, n_0[i], n_1[i],   n_2[i], k_2[i], d_2, n_3[i], k_3[i], d_3)-R_{exp}}{\sigma_R}\right)^2
        \label{eq:chi2}
    \end{equation}
    where $i$ and $j$ are the wavelength indices ($\lambda_i$) and AOIs ($\alpha_j$), respectively; and $R_{theo}$ is the theoretically predicted value based on Eq.~\ref{eq:TMM_Incoherent}. $R_{exp}$ is the reflectance from experimental data, and $\sigma_R$ is the error of $R_{exp}$. $n_0$ and $n_1$ are the refractive indices of LAB and PMT glass that are fixed in the fitting and shown in Figure~\ref{fig:nk_glass_lab}. The refractive indices of ARC and photocathode are complex numbers: $\Tilde{n}_2=n_2+ik_2$ and $\Tilde{n}_3=n_3+ik_3$, and $d_2$ and $d_3$ are their respective thicknesses. $n_{2(3)}$, $k_{2(3)}$ and $d_{2(3)}$ are free parameters in the fitting.
   
\begin{figure}
        \centering
        \subfloat[Refractive indices of ARC]{
        \label{fig:fig4a}
        \includegraphics[scale=0.55]{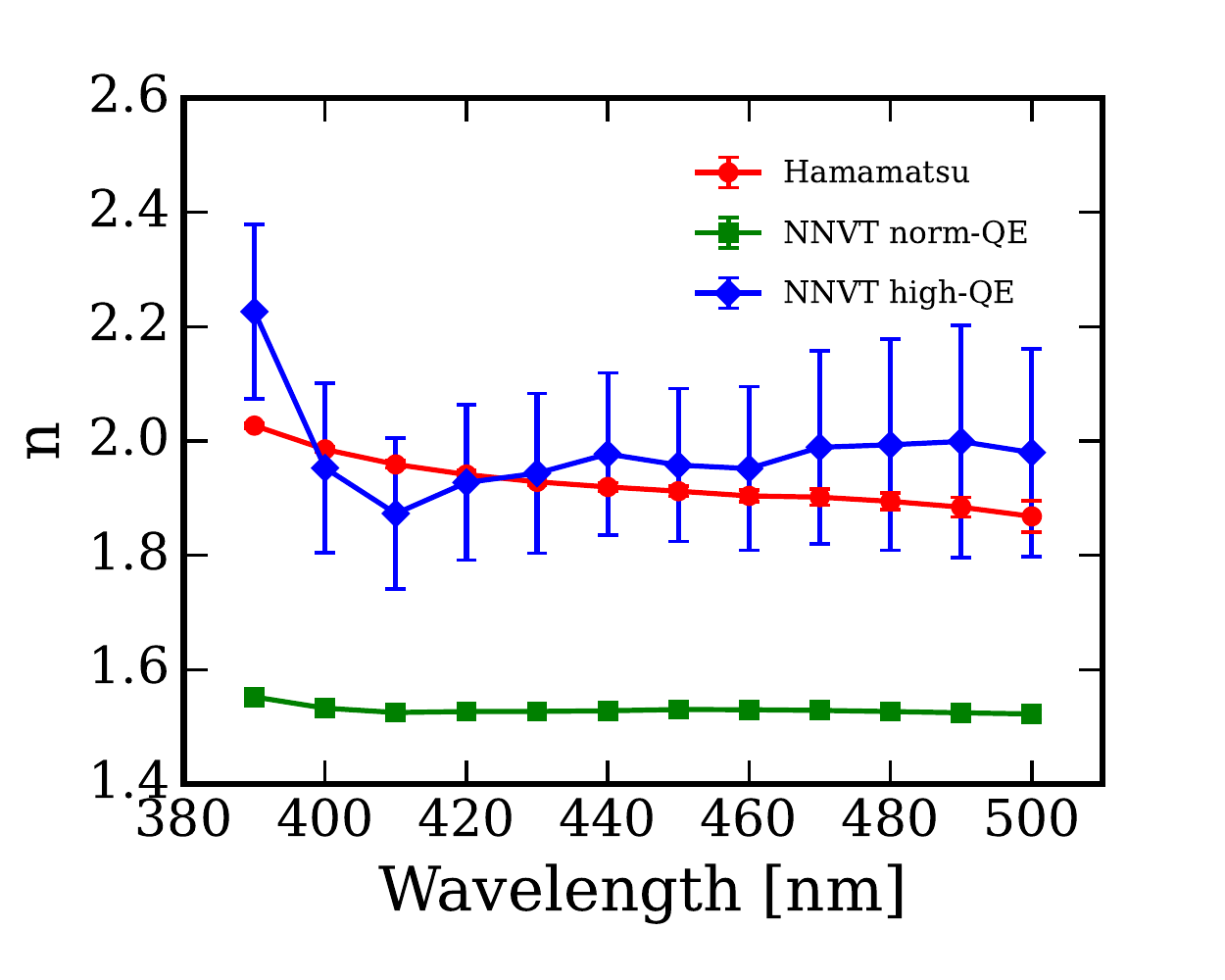}
        }
        \subfloat[Refractive indices and extinction coefficients of photocathode]{
        \label{fig:fig4b}
        \includegraphics[scale=0.55]{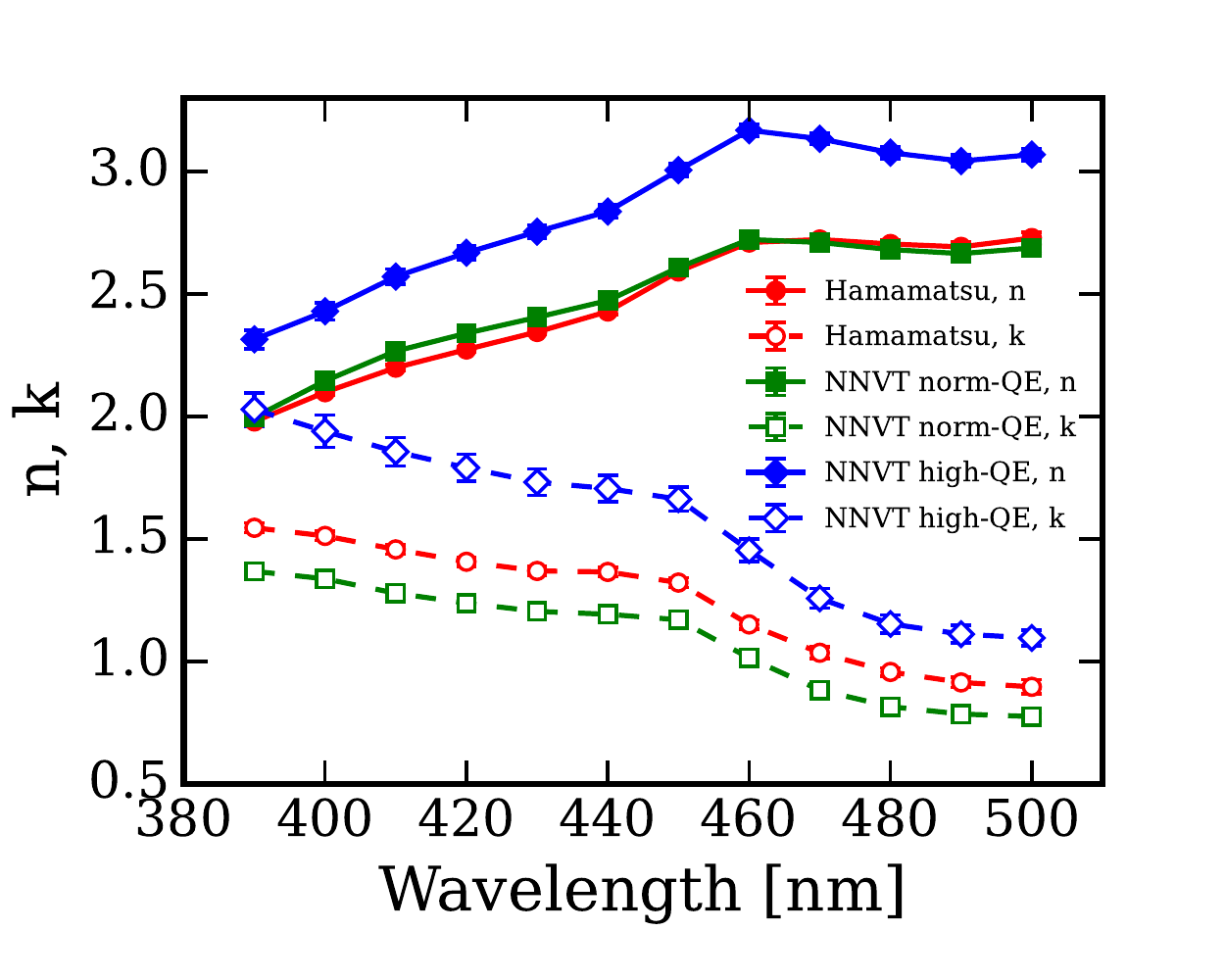}
        }
        \caption{Spectral dependence of refractive indices of ARC (a) and refractive indices (solid lines) and extinction coefficients (dashed lines) of photocathode (b) for the three PMTs of R12860 (red), high-QE MCP-PMT (green) and normal-QE MCP-PMT (blue)}
        \label{fig:fig4}
    \end{figure}

    As an example, Figure~\ref{fig:fig3} shows the fitted curves that agree with the reflectance data points for the three 20'' PMTs at four wavelengths. We obtain a similar fitting quality at all other measured wavelengths. Figure~\ref{fig:fig4} shows the fitting results of the refractive indices and extinction coefficients as a function of wavelength, and Table \ref{tab:tab1} summarizes the fitting results of thicknesses of the ARC and photocathode of the three PMTs. The absorption of ARC is found to be negligible for the three PMTs in the range of measured wavelengths, and its extinction coefficients are set to 0. Figure \ref{fig:fig4a} shows that the normal-QE MCP-PMT uses different ARC materials compared with the other two tubs due to the significant difference in ARC optical properties. For the high-QE MCP-PMT, the figure shows much larger uncertainties in the refractive index of ARC, which is caused by the PMT's small thickness of the ARC layer. Thus, in this case, the $\chi^2$ values are not sensitive to the variations of ARC's refractive index. Figure~\ref{fig:fig4b} shows the refractive indices (solid lines) and extinction coefficients (dashed line) of the photocathode as a function of wavelength. Even though the three PMTs use the same type of photocathode material (i.e. a bialkali photocathode), their optical parameters are markedly different; this result might be primarily caused by different processes and technologies during photocathode manufacturing.
    
    \begin{table}
        \centering
        \begin{tabular}{|c|c|c|c|}
        \hline
        \diagbox{Layer}{PMT type} & Hamamatsu & NNVT high-QE & NNVT normal-QE \\
        \hline
        ARC & $36.5\pm0.4$ & $10.2\pm1.2$ & $49.1\pm5.2$ \\
        \hline
        Photocathode & $21.1\pm0.2$ & $18.7\pm0.4$ & $20.4\pm0.3$ \\
        \hline
        \end{tabular}
        \caption{Thicknesses of ARC and photocathode of the three 20'' PMTs, in unit of nm}
        \label{tab:tab1}
    \end{table}
   
    \subsection{Other optical parameters in the PMT optical model}
\begin{figure}
        \centering
        \subfloat[Hamamatsu PMT]{
        \label{fig:fig5a}
        \includegraphics[scale=0.35]{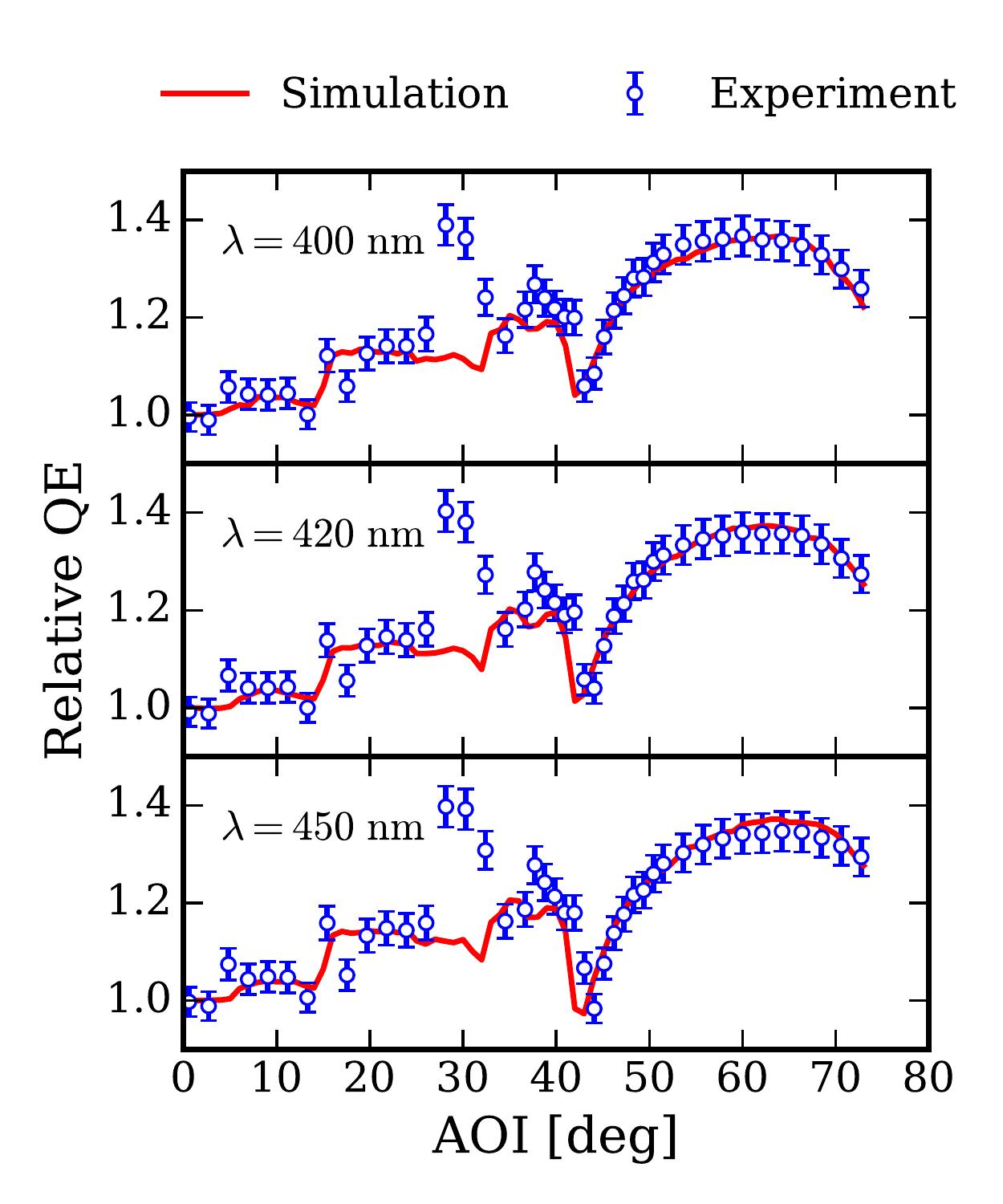}
        }
        \subfloat[NNVT high-QE PMT]{
        \label{fig:fig5b}
        \includegraphics[scale=0.35]{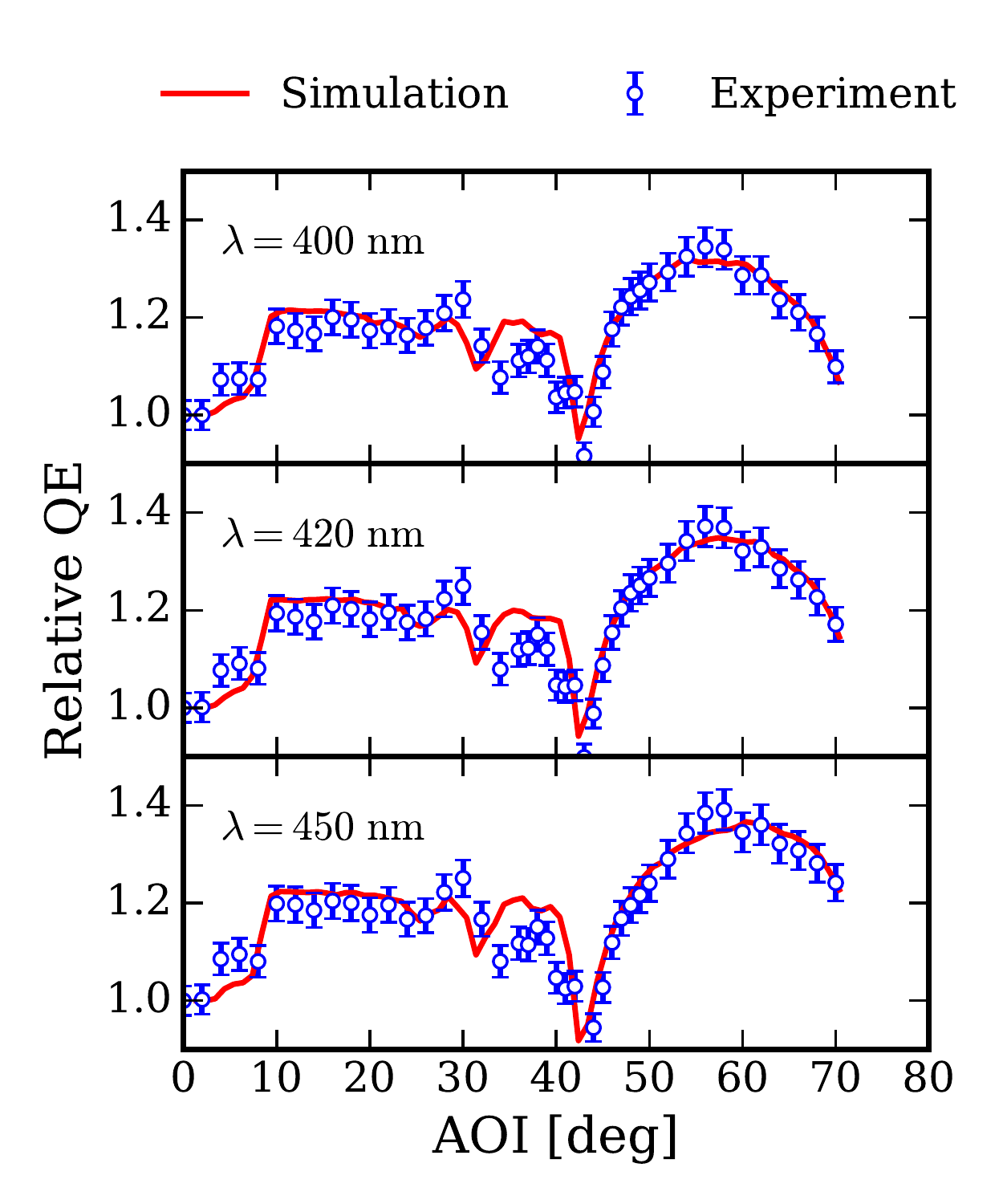}}
        \subfloat[NNVT normal-QE PMT]{
        \label{fig:fig5c}
        \includegraphics[scale=0.35]{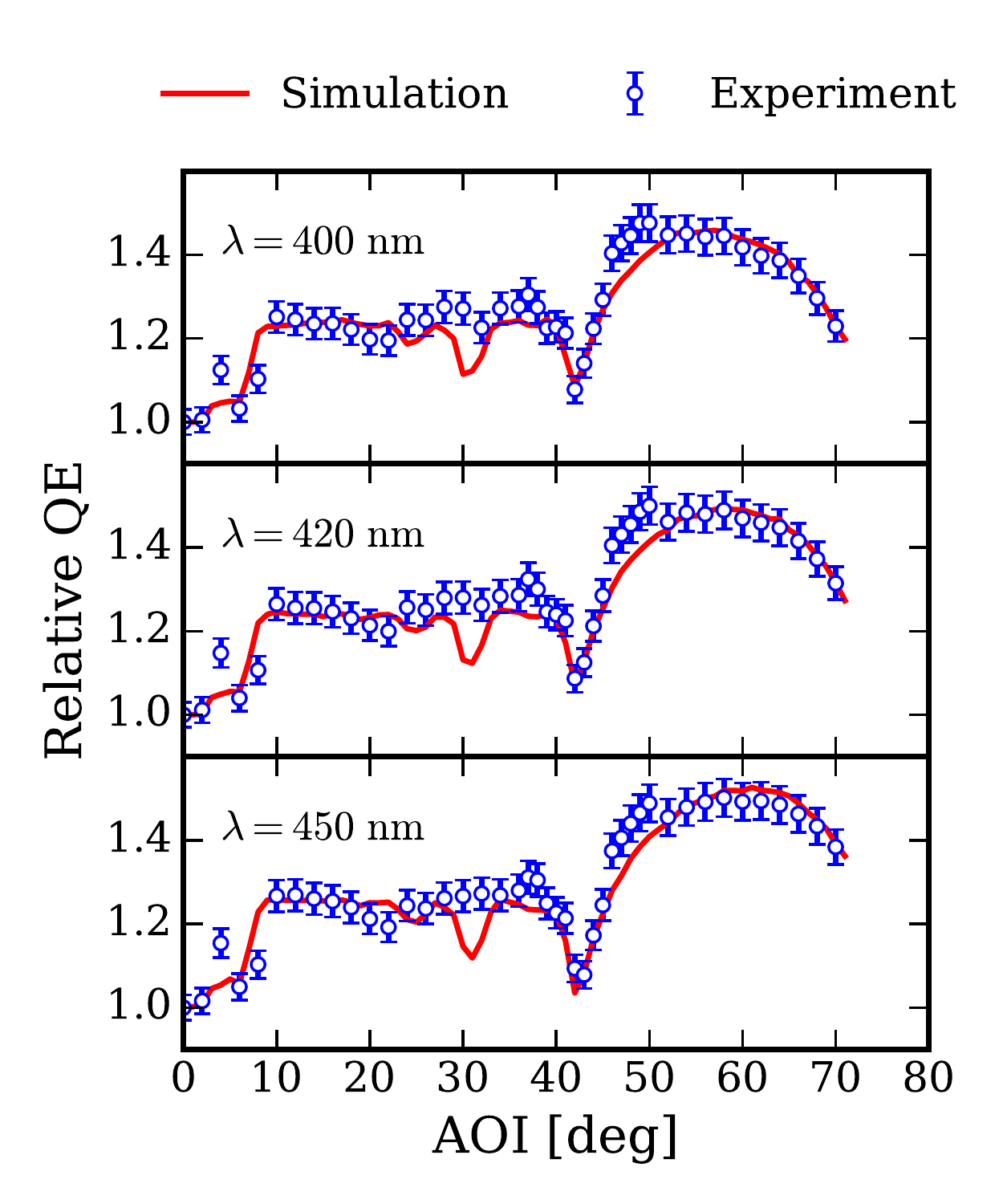}}
        \caption{Relative QE as a function of AOI measured at three typical wavelengths for the Hamamatsu PMT (a), the NNVT high-QE PMT (b) and the NNVT normal-QE PMT (c). The blue markers denote the experimental data, and the red curves represent the prediction from the PMT optical model.}
        \label{fig:qe_response}
    \end{figure}

    The remaining parameters in the PMT optical model, including the optical properties of the inner structures and escape factor of the photocathode, can be constrained appropriately by the angular responses of the PMTs' QE measured at different wavelengths in the LAB. Figure~\ref{fig:qe_response} shows a few examples of relative QE as a function of AOI measured at wavelengths of 400~nm (top), 420~nm (middle) and 450~nm (bottom) with blue markers for the Hamamatsu R12860 (Figure~\ref{fig:fig5a}), the high-QE MCP-PMT (Figure~\ref{fig:fig5b}) and the normal-QE MCP-PMT (Figure~\ref{fig:fig5c}). The QE is normalized to an AOI of 0. In the region of total internal reflection (i.e., larger than 42 degrees), the QE increases and shows a smooth structure due to more light absorption in the photocathode and no light transmitted into the interior of the PMT; therefore, no influences from the inner structures are observed. The erratic fluctuations between 0 and 42 degrees are caused by the optical processes occurring inside the PMT. When AOI is near 0, the transmitted light hits the dynode or MCP that has negligible reflectance directly, and the QE at approximately 0 degrees is small due to no contributions from other positions on the photocathode. As the AOI gradually increases, the transmitted light begins to be reflected by the inner electrodes or the aluminum film in the lower hemisphere or even directly hit other positions of the photocathode. Different optical processes lead to different contributions to the QE amplitude.
    
\begin{figure}
        \centering
        \includegraphics[width=300px]{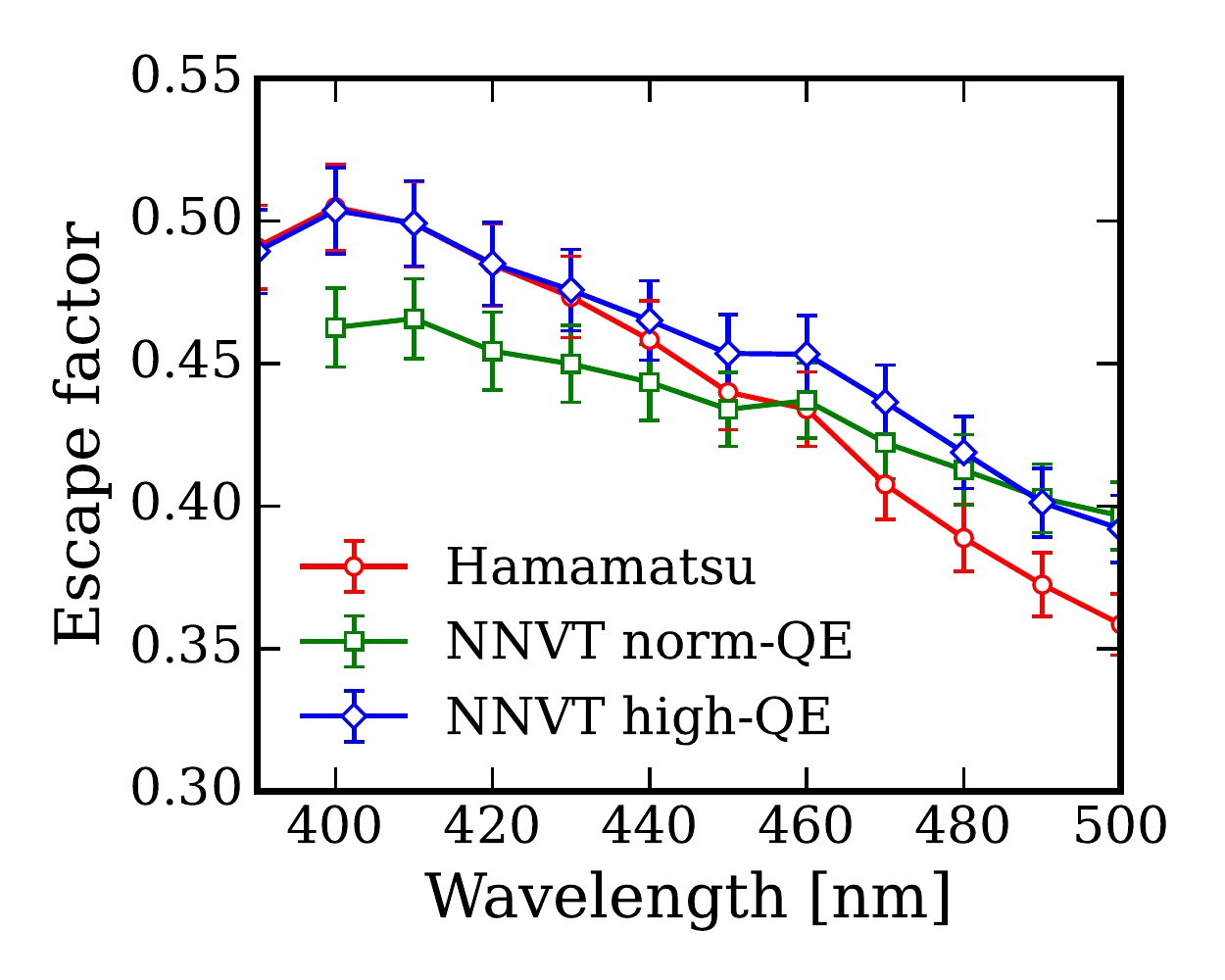}
        \caption{Escape factor as a function of wavelength for R12860 from Hamamatsu (red), normal-QE MCP-PMT (green) and high-QE MCP-PMT (blue) from NNVT.}
        \label{fig:escape_fac}
    \end{figure}

    In the constructed geometry models, as discussed in section~\ref{pmtgeometry}, we set the reflectivity of the aluminum film to 92\%~\cite{Rakic:95} and tune the reflectivity of the cylindrical tube made of stainless steel to be 65\% for all PMTs, the same reflectivity is applied for electrodes of NNVT PMTs; while for electrode of Hamamatsu PMT, it is set to 20\% due to some stains on its surface which might be caused by electric welding.  The escape factor as a function of wavelength is shown in Figure~\ref{fig:escape_fac}. Then, the new PMT optical model can well describe the angular responses of QE for the three PMTs, which are shown as red lines in Figure~\ref{fig:qe_response}. Even though there are still a few inconsistencies between the QE data and the model predictions, most can be suppressed by the average effect when we consider a photodetector system that consists of tens of thousands of PMTs. Also, agreement can be improved by considering more detailed structures in the geometry model. For example, the QE from the experimental data is much larger than that from the simulation at approximately 30 degrees for all three PMTs, which is caused by an overlap between the photocathode and the aluminum film around the PMT equator. In this case, the refracted light hits on the overlapped region and might generate photoelectrons. Meanwhile, a fraction of the refracted light could pass through the photocathode and be reflected by the aluminum film, which has also a chance to generate photoelectrons in the photocathode. For both cases, the generated photoelectrons could contribute to the QE. Because the structure of the overlapped region strongly depends on the technologies used in PMT manufacturing, it could be significantly different among different types of PMTs. However, the impacts on PDE from this inconsistency can be mitigated by the low CE in this region. Similar agreements are also achieved at other measured wavelengths.
    
    \section{Light yield excess in a liquid scintillator detector}
    The light yield excess is observed in a few past neutrino experiments. In general, the light yield from the experimental data is 20\%-30\% higher than the predictions at the R\&D phases. They are summarized below:

    \begin{enumerate}[(a)]
        \item Borexino: The predicted light yield is 400~p.e./MeV in its design report~\mbox{\cite{ALIMONTI2002205}} and the measured one is 500~p.e./MeV~\mbox{\cite{Borexino:2008gab}}, corresponding to 25\% light yield excess.
        \item Daya Bay: The predicted light yield is 105~p.e./MeV in its technical design report~\mbox{\cite{DayaBay:2007fgu}} and updated to 134~p.e./MeV in 2009. The measured number is 162~p.e./MeV~\mbox{\cite{DayaBay:2012fng}}. So, the excess is about 21\%. In its prototype detector, the light yield is 200~p.e./MeV from the simulation, compared with 240~p.e./MeV from the experimental data, which yields an excess of 20\%~\mbox{\cite{DayaBay:2007fgu}}.
        \item KamLAND: The predicted light yield is 150~p.e./MeV presented in Neutrino 2000 conference~\mbox{\cite{Piepke:2001tg}} and the measured one is $\sim$280~p.e./MeV~\mbox{\cite{KamLAND:2002uet}}, which leads to more than 80\% light yield excess. However, based on~\mbox{\cite{suekane2004overview}}, a large fraction of the excess is caused by the light re-emission and scattering that are not considered in the predicted light yield.
        \item RENO: The predicted energy resolution is $6.5\%/\sqrt{E}$ stated in its design report~\mbox{\cite{RENO:2010vlj}} and the measured energy resolution is $5.9\%/\sqrt{E}$~\mbox{\cite{RENO:2012mkc}}. Since the energy resolution is dominated by photoelectron statistics, the light yield excess is estimated to be about 21\%.
    \end{enumerate}

    A toy Monte Carlo study is performed to evaluate the impact of the new PMT optical model on the light yield of a simple liquid scintillator detector, compared with that from the simplified PMT optical model, which assumes that the light is 100\% absorbed and converted to free photoelectrons by simply applying PMT's QE. The new model considers the PDE angular response, reflection on the photocathode and optical processes inside the PMT, and all these effects predict a higher light yield than the simplified model. However, because the optical processes inside the PMT are strongly dependent on the PMT geometry, we only investigated the former two effects in this study. 
    
\begin{figure}
        \centering
        \subfloat[LS emission spectrum]{
        \label{fig:fig6a}
        \includegraphics[scale=0.5]{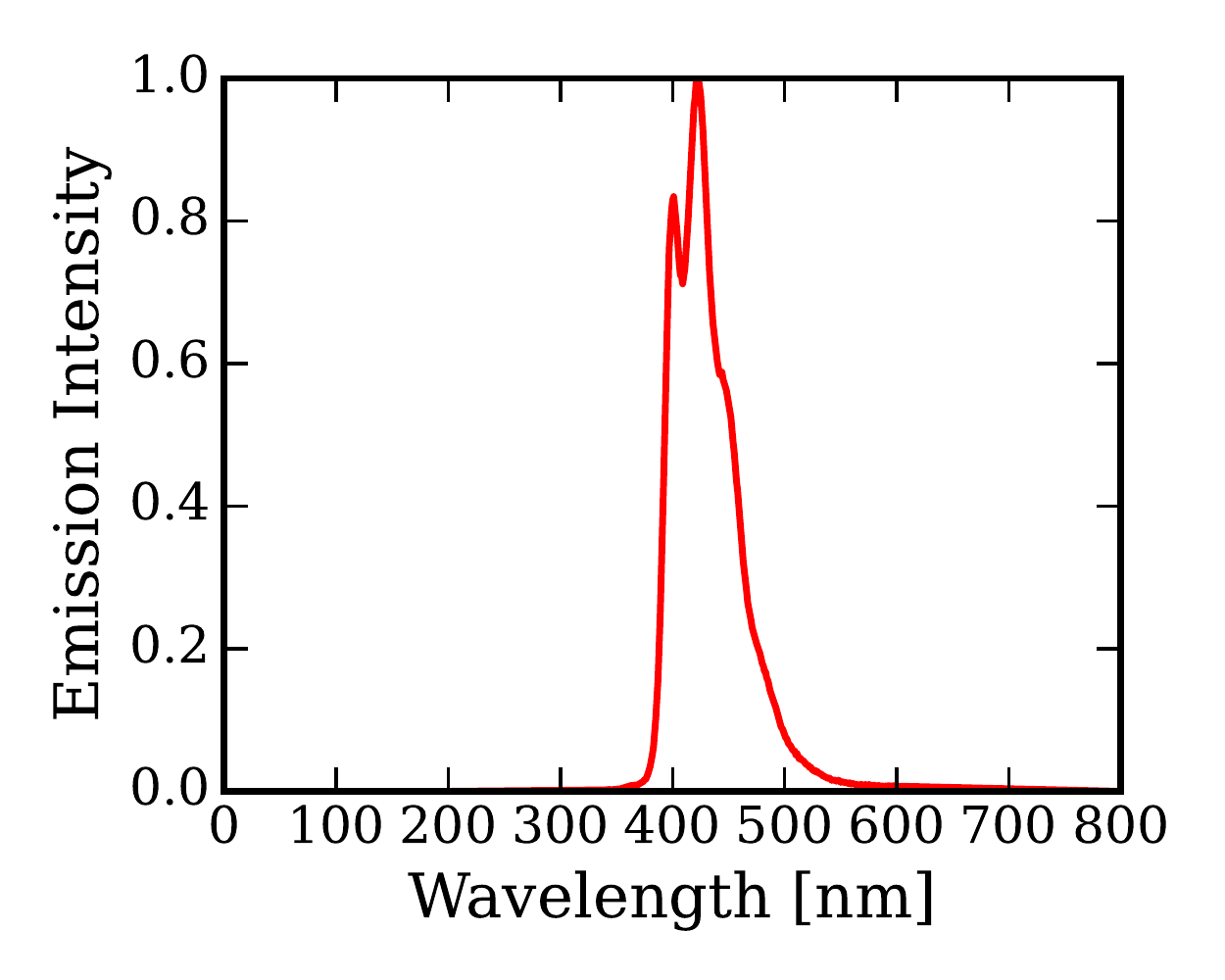}}
        \subfloat[AOI distribution at photocathode]{
        \label{fig:fig6b}
        \includegraphics[scale=0.5]{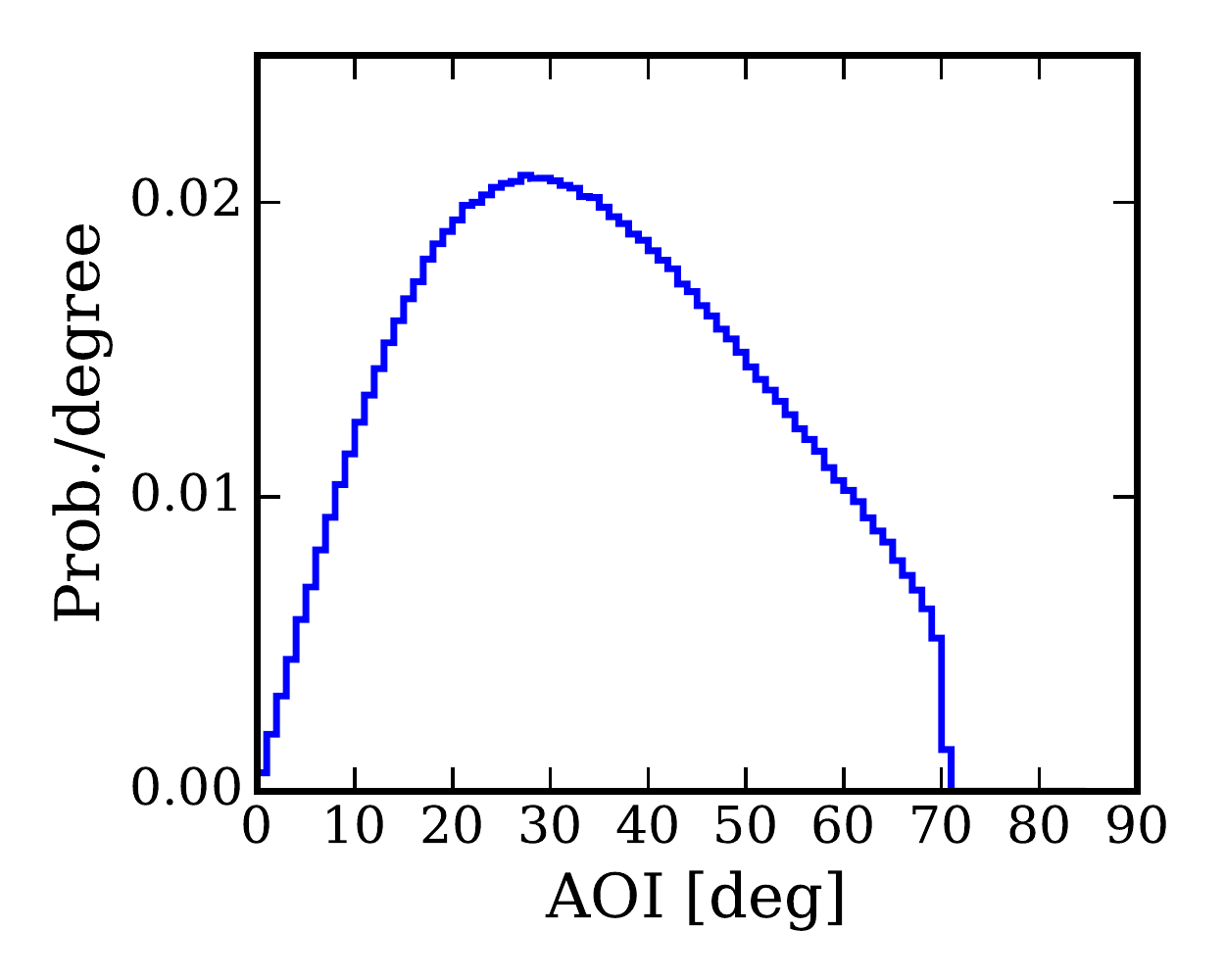}}

        \subfloat[Typical QE curve]{
        \label{fig:fig6c}
        \includegraphics[scale=0.5]{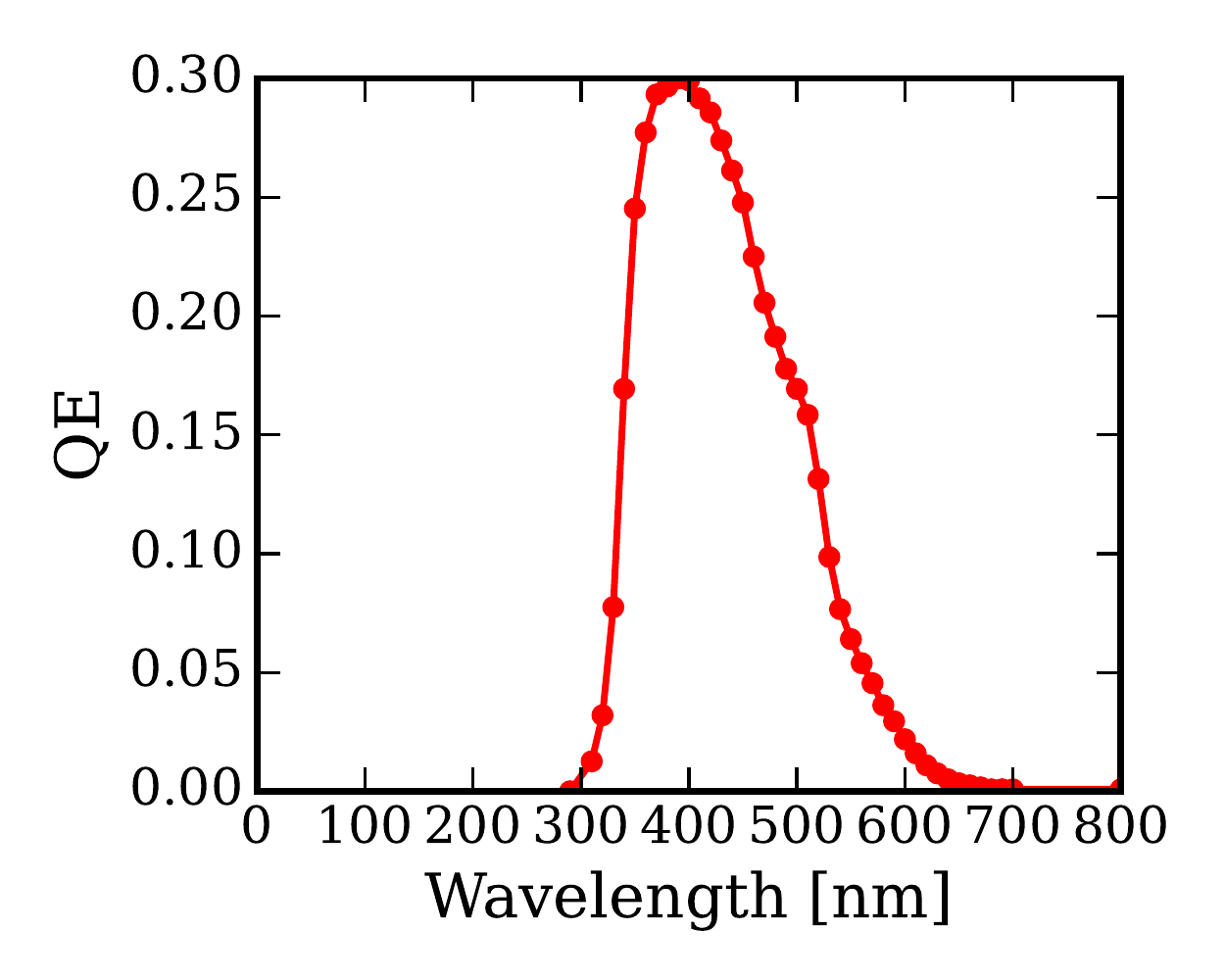}}
        \subfloat[Reflectance and absorbance at 420~nm]{
        \label{fig:fig6d}
        \includegraphics[scale=0.5]{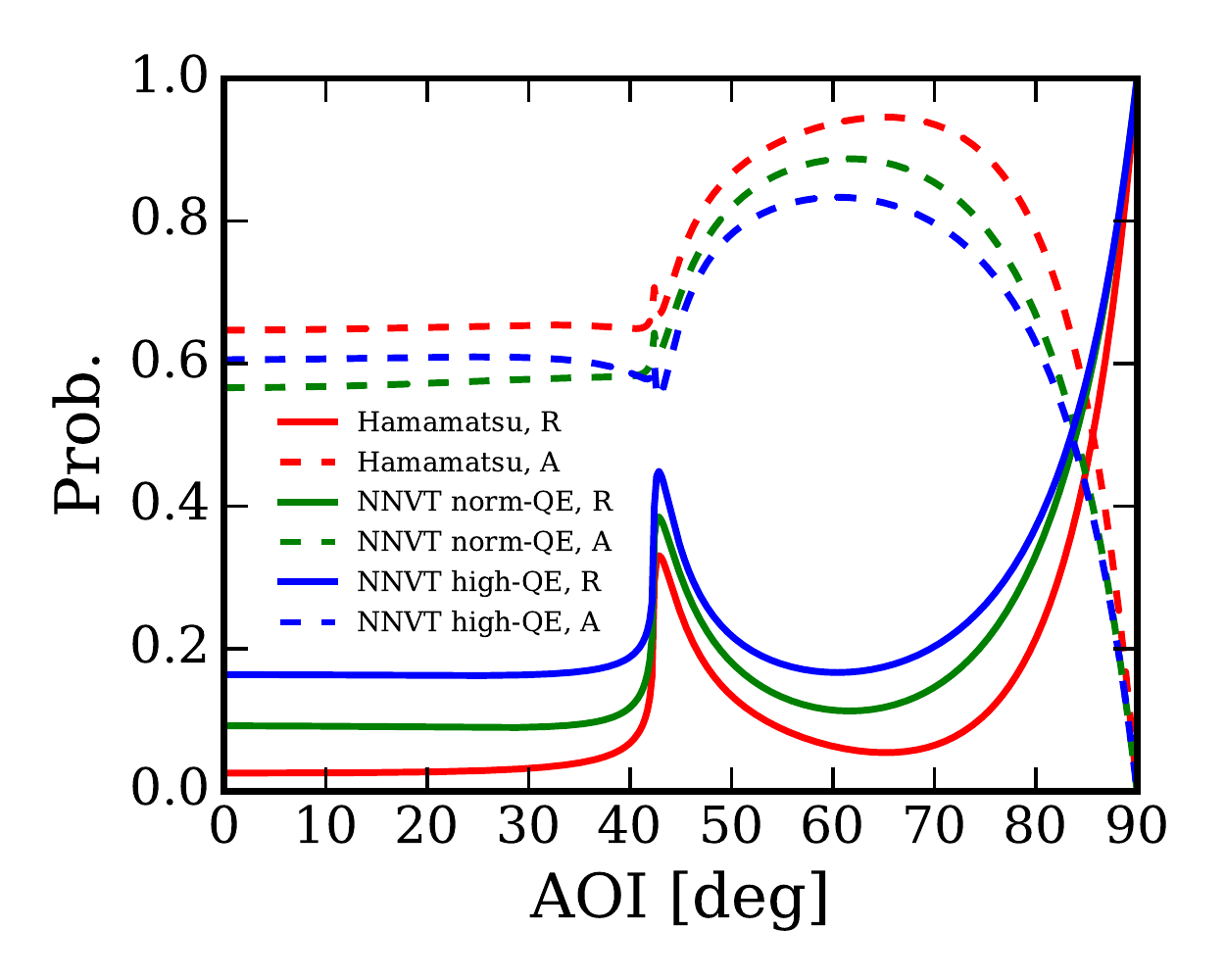}}
        \caption{Input parameters of the toy Monte Carlo simulation}
        \label{fig:fig6}
    \end{figure}

    In the new model, both the PMT angular response and reflection are spectrally and angularly dependent. In the toy Monte Carlo study, the wavelength is sampled based on the emission spectrum of Daya Bay's liquid scintillator (Figure~\ref{fig:fig6a})~\cite{Beriguete:2014gua}. For the AOI distribution at the photocathode, we assume a uniform and parallel light beam illuminated on the entire PMT window (perpendicular to the plane of PMT's equator) and the PMT's working medium to be water. Then, the AOI distribution is obtained and shown in Figure~\ref{fig:fig6b}, which terminates at approximately 70 degrees, because of the light refraction between water and glass. A typical spectral response of QE is shown in Figure \ref{fig:fig6c} for a R12860 PMT from Hamamatsu~\cite{datasheet}. With the optical parameters determined in this study, the reflectance and absorbance are calculated for the three 20'' PMTs at wavelengths of interest. Figure~\ref{fig:fig6d} shows an example of reflectance (solid) and absorbance (dashed) curves at 420~nm. Regarding reflection on PMTs, its impact on the light yield depends on the photocathode coverage. We assume 100\% coverage in this study, and its impact should be scaled with the real coverage for a certain detector. 
    
    Table \ref{tab:tab2} summarizes the fraction of light yield excess from different factors in the new PMT model, compared with light yield obtained with the simplified PMT model. When the PMTs are immersed in liquid scintillator, both the reflectance and absorbance of photocathode are enhanced, because of effects of the total internal reflections. Since the new model can reasonably manage these effects, it indicates a significant contribution to the light yield in a liquid scintillator detector from these two effects. In general, the new model predicts a $20\% \sim 30\%$ increase in light yield compared with the simplified one, which is consistent with the experimental observations. However, this estimation is relatively rough because the result should also be correlated with the properties of the liquid scintillator, such as its attenuation length and re-emission probability. Geometries of liquid scintillator detectors may also affect the AOI distribution at the photocathode. A full Monte Carlo study is recommended to describe a specific liquid scintillator detector.
    
    \begin{table}
        \centering
        \begin{tabular}{|c|c|c|c|}
        \hline
        \diagbox{Effect}{PMT type} & Hamamatsu & NNVT normal-QE & NNVT high-QE \\
        \hline
        PDE angular response & 11.8\% & 16.0\% & 8.9\% \\
        \hline
        Reflection & 9.5\% & 14.0\% & 20.3\% \\
        \hline
        Total & 21.3\% & 30.0\% & 29.2\% \\
        \hline
        \end{tabular}
        \caption{Contribution of PDE angular response and reflection on photocathode to light yield}
        \label{tab:tab2}
    \end{table}
    
    \section{Conclusion}
    A reliable and accurate PMT optical model is critical to correctly describe the angular and spectral responses of the PDE in a PMT working medium in many applications. In this study, we proposed a new PMT optical model to manage both light interactions with the PMT window and optical processes occurring inside PMTs. Due to the internal optical processes, the PDE could be contributed to by multiple other positions on the photocathode in addition to the incident position. The new model builds a relationship between the PDE and the underlying processes that the PDE relies on, including light absorption of photocathode, escape probability of photoelectrons, collection efficiency and light transportation inside the PMT. The light absorption coefficient of photocathode can be calculated based on the optics theory and the known optical parameters of ARC, photocathode, PMT glass and external media. The escape probability of photoelectrons and collection efficiency are considered as a product (called $F$ factor) in the model. The $F$ factor at one position on the photocathode can be interpolated with the values of $F$ at a few reference positions for the case of non-uniform photocathode. The optical processes inside the PMT determine the number of positions contributed to the PDE and their weight coefficients. They can be reasonably handled by the Monte Carlo simulation with a detailed PMT geometry model. Both the $F$ factors and weight coefficients can be determined together by combining the information of the PDE measured at these reference positions and Monte Carlo simulation of the optical processes inside the PMT. A complete procedure has been established to demonstrate the methods used to obtain the key optical parameters in the model using the two 20'' MCP-PMTs from NNVT and one 20'' dynode PMT from Hamamatsu. In this process, the refractive indices, extinction coefficients and thicknesses of the ARC and photocathode are extracted by fitting the angular response of reflectance in LAB at wavelengths between 390 and 500~nm for the three 20'' PMTs. Fitted results show that the optical properties of the ARC and photocathode of the three PMTs are significantly different, which indicates that they use different ARC materials or technologies during PMT fabrication. The remaining key parameters in the model, such as the reflectivity of inner structures and escape factors, are well constrained by the collected QE data. Results show that the new PMT model can reproduce the angular response of QE with good precision for most angles, even for fine structures in the spectra. The discrepancy around 30 degrees is caused by the overlap region between the photocathode and the aluminum film, nevertheless, its impacts on PDE can be mitigated due to low CE in this region.
    
    We also performed a toy Monte Carlo study to estimate the light yield of a liquid scintillator detector with a simple configuration. Using the parameters obtained in this study, the new PMT optical model predicts 20\% to 30\% more light yield than the simplified PMT model, which assumes that the light hit on PMTs is 100\% absorbed and converted to free photoelectrons by simply applying PMT's QE. The level of light yield excess is similar to that observed in various liquid scintillator-based neutrino detectors. These results imply that the fundamental reason for this excess is caused by imperfect PMT optical models in their detector simulations. 
    
    In most applications, PMTs are characterized in air, however, they may eventually be operated in other media. As discussed above, the PDE evaluated in air does not represent the PMT's PDE in other external media. The proposed new PMT model can naturally convert the PDE from one medium to another by simply changing the optical parameters of external media. Finally, we demonstrated the performance of the new PMT model using three 20'' PMTs and relevant characterizations performed in laboratory experiments. However, more studies will be required to validate this model for other types of PMTs, and more efforts are required to build a model for large photodetector systems in which the variations among PMTs must be considered.

    \section{Acknowledgments}
    We gratefully acknowledge support from National Natural Science Foundation of China (NSFC) under grant No. 11875279. This work is also supported in part by the Strategic Priority Research Program of the Chinese Academy of Sciences, Grant No. XDA10010800, and the CAS Center for Excellence in Particle Physics (CCEPP). We are immensely grateful to Prof. Dmitry Naumov and Dr. Tatiana Antoshkina from JINR for discussions on the part of optics theory. We are also grateful to Yuduo Guan, Zhenning Qu and Chengzhuo Yuan for assistance with equipment improvement and data taking.
  
  %  \appendix
  % \section{}
  %  \begin{table}[h]
  %      \centering
  %      \begin{tabular}{|c|c|c|c|c|c|c|}
  %      \hline
  %      \multirow{2}*{$\lambda$ [nm]} & \multicolumn{2}{c|}{Hamamatsu} &\multicolumn{2}{c|}{NNVT normal-QE} & \multicolumn{2}{c|}{NNVT high-QE} \\
  %      \cline{2-7}
  %      ~ & ARC & Pho. & ARC & Pho. & ARC & Pho. \\
  %      \hline
  %      390 & 2.03 & 1.98+1.55i & 1.55 & 2.00+1.37i & 2.23 & 2.31+2.03i \\
  %      400 & 1.99 & 2.10+1.51i & 1.53 & 2.15+1.34i & 1.95 & 2.43+1.94i \\
  %      410 & 1.96 & 2.20+1.46i & 1.53 & 2.27+1.28i & 1.87 & 2.57+1.86i \\
  %      420 & 1.94 & 2.27+1.41i & 1.53 & 2.34+1.24i & 1.93 & 2.67+1.79i \\
  %      430 & 1.93 & 2.35+1.37i & 1.53 & 2.40+1.20i & 1.94 & 2.75+1.73i \\
  %      440 & 1.92 & 2.43+1.37i & 1.53 & 2.47+1.19i & 1.98 & 2.84+1.71i \\
  %      450 & 1.91 & 2.59+1.32i & 1.53 & 2.61+1.17i & 1.96 & 3.01+1.66i \\
  %      460 & 1.90 & 2.71+1.15i & 1.53 & 2.72+1.01i & 1.95 & 3.17+1.45i \\
  %      470 & 1.90 & 2.72+1.04i & 1.53 & 2.71+0.88i & 1.99 & 3.13+1.26i \\
  %      480 & 1.89 & 2.70+0.96i & 1.53 & 2.68+0.82i & 1.99 & 3.08+1.15i \\
  %      490 & 1.88 & 2.69+0.91i & 1.52 & 2.66+0.79i & 2.00 & 3.04+1.11i \\
  %      500 & 1.87 & 2.73+0.90i & 1.52 & 2.69+0.78i & 1.98 & 3.07+1.10i \\
  %      \hline
  %      \end{tabular}
  %      \caption{Best fit values of the refractive indices of ARC and photocathode for the three types of PMTs}
   %     \label{tab:tab3}
    %\end{table}

\printbibliography
% \bibliographystyle{elsarticle-num}
% \bibliography{references}

\end{document}